\documentclass[pra,showpacs,twocolumn,aps,superscriptaddress,a4paper,longbibliography]{revtex4-1}
\usepackage{dcolumn,amssymb,amsmath,amsfonts,graphicx,latexsym,color}
\usepackage[utf8]{inputenc}
\usepackage[T1]{fontenc}
\usepackage{epstopdf}

\usepackage{ulem}
\usepackage{epstopdf}
\usepackage{subfigure}
\usepackage{float}
\usepackage{mathrsfs}
\usepackage{bbold}
\usepackage{psfrag}

\usepackage{verbatim}
\usepackage{multirow}
\usepackage{diagbox}
\usepackage[colorlinks,citecolor=blue]{hyperref}

\def\CH{\textcolor{red}}

\begin{document}


\title{Light-induced phase crossovers in a quantum spin Hall system}

\author{Fang Qin}
\email{qinfang@nus.edu.sg}
\affiliation{Department of Physics, National University of Singapore, Singapore 117542, Singapore}

\author{Ching Hua Lee}
\email{phylch@nus.edu.sg}
\affiliation{Department of Physics, National University of Singapore, Singapore 117542, Singapore}

\author{Rui Chen}
\email{chenr@hubu.edu.cn}
\affiliation{Department of Physics, Hubei University, Wuhan 430062, China}
\affiliation{Department of Physics, The University of Hong Kong, Pokfulam Road, Hong Kong 999077, China}

\date{\today }

\begin{abstract}
In this work, we theoretically investigate the light-induced topological phases and finite-size crossovers in a paradigmatic quantum spin Hall (QSH) system with high-frequency pumping optics. Taking the HgTe quantum well for an example, our numerical results show that circularly polarized light can break time-reversal symmetry and induce the quantum anomalous Hall (QAH) phase. In particular, the coupling between the edge states is spin dependent and is related not only to the size of the system, but also to the strength of the polarized pumping optics. By tuning the two parameters (system width and optical pumping strength), we obtain four transport regimes, namely, QSH, QAH, edge conducting, and normal insulator. These four different transport regimes have contrasting edge conducting properties, which will feature prominently in transport experiments on various topological materials. 
\end{abstract}

\maketitle

\section{Introduction}\label{1}

The quantum spin Hall (QSH) insulator is a member of the family of quantum Hall insulators first proposed by Kane and Mele~\cite{kane2005z,kane2005quantum}. 
It is an energy band insulator in bulk, but conducting at the edges~\cite{abergel2018hot}. The edges carry two different charge currents with spin-up and spin-down moving in opposite directions.
A more realistic version of the QSH effect in HgTe quantum wells was proposed by Bernevig, Hughes, and Zhang~\cite{bernevig2006quantum} and was soon confirmed by experiment~\cite{konig2007quantum,konig2008quantum}.
Since then, there have been intensive studies to investigate the exotic properties in QSH systems~\cite{prodan2009robustness}. They include topological field theory~\cite{qi2008topological,yu2011equivalent,lee2015free}, finite-size crossovers in topological insulators~\cite{zhou2008finite,liu2010oscillatory}, magnetic doping in QSH system (Mn-doped HgTe quantum well)~\cite{fu2014finite}, disorder-induced topological Anderson insulator in QSH systems~\cite{li2009topological,jiang2009numerical,groth2009theory,xing2011topological,li2011size,ning2022photoinduced}, Floquet topological insulator states in QSH systems~\cite{calvo2015floquet,zhu2014floquet,qin2022phase,liu2021floquet}, as well as QSH effects in \CH{} three-dimensional topological insulators~\cite{li2010chern,dabiri2021light,dabiri2021engineering,pervishko2018impact,zhu2022floquet}.

The quantum anomalous Hall (QAH) insulator, which was first proposed by Haldane~\cite{haldane1988model}, is another key member of the family of quantum Hall insulators. A QAH insulator can have a quantized Hall conductivity even without an external magnetic field that breaks the time-reversal symmetry~\cite{haldane1988model,chang2013experimental,yu2010quantized,linder2009anomalous,lu2013quantum,lu2010massive,shan2010effective}. The QAH effect was first experimentally observed in Ref.~\cite{chang2013experimental}; 
in another work on finite-size effects in QAH systems~\cite{fu2014finite}, they adopted the magnetic doping (Mn doping) to break the time-reversal symmetry and introduce the QAH phase in the HgTe quantum well. By appropriate tuning of the doping concentration and the system size, they have obtained a variety of topological phases~\cite{fu2014finite}. 

Yet, this experimental success~\cite{fu2014finite} in realizing QAH states proved difficult to follow up since the doping concentration is difficult to tune continuously in experiments. Therefore, departing from previous works~\cite{fu2014finite}, we shall provide a more experimentally accessible alternative to realizing QAH states, using circularly polarized light instead of magnetic doping to break the time-reversal symmetry in HgTe quantum well. This hinges on the fact that the intensity of circularly polarized light can be easily tuned continuously in experiments, and that Floquet driving can induce a host of exotic phases in various media, such as disorder-induced Floquet topological insulators~\cite{titum2015disorder}, quenched topological boundary modes induced by Floquet engineering~\cite{lee2021quenched}, Floquet semimetal with exotic topological linkages~\cite{li2018realistic}, Floquet mechanism for non-Abelian fractional quantum Hall states~\cite{lee2018floquet}, photoinduced half-integer quantized conductance plateaus in topological insulators~\cite{yap2018photoinduced}, and tunable Floquet-Weyl semimetals driven from nodal line semimetals~\cite{chan2016type,bonasera2022tunable}. 
In this work, based on numerical calculations with high-frequency Floquet expansion, we find that circularly polarized light can break the time-reversal symmetry and introduce the QAH phase in a HgTe quantum well. Our findings complement other works that utilize circularly polarized light to induce new topological phases in graphene~\cite{oka2009photovoltaic,sato2019microscopic,mciver2020light,sentef2015theory,broers2022detecting,candussio2022terahertz,qin2022nondiagonal,cupo2021floquet,assi2021floquet,luo2021tuning,schuler2020circular,topp2019topological,rodriguez2020floquet,li2020floquet,berdakin2021spin,dal2017one,huaman2021quantum}, but with significant differences. In particular, the coupling between the edge states is spin dependent and is related not only to the width of the system, but also to the strength of the polarized pumping optics. By appropriate tuning of the system size and strength of the pumping optics, we obtain four transport regimes including QSH, QAH, edge conducting (EC), and normal insulator (NI). 

The paper is organized as follows. In Section~\ref{2}, we introduce the model Hamiltonian that we use. In Section~\ref{3}, we derive the corresponding Floquet effective Hamiltonian by using the high-frequency expansions. In Section~\ref{4}, we calculate the energy dispersions under the open boundary condition along the $y$ direction and periodic boundary conditions along the $x$ direction with fixed light intensity and different system size by numerically diagonalizing the corresponding tight-binding Hamiltonian. In Section~\ref{5}, we give and discuss our phase diagram. 
Finally, we summarize our results in Section~\ref{6}.

\section{Model}\label{2}

We start with the Bernevig-Hughes-Zhang (BHZ) model given by the low-energy two-dimensional effective Hamiltonian near the $\Gamma$ point to describe the electronic states in HgTe/(Hg,Cd)Te quantum wells~\cite{bernevig2006quantum,konig2007quantum,konig2008quantum,zhou2008finite,rothe2010fingerprint,zhu2014floquet}
\begin{align}\label{eq:H_BHZ}
{\cal H}({\bf k}) &=
\begin{pmatrix}
    h({\bf k}) & 0 \\
    0 & h^{*}(-{\bf k})
\end{pmatrix} \nonumber\\
&=\epsilon_{\bf k}\tau_{0}\otimes\sigma_{0} + Ak_{x}\tau_{z}\otimes\sigma_{x} + Ak_{y}\tau_{0}\otimes\sigma_{y} \nonumber\\
&+ {\cal M}_{\bf k}\tau_{0}\otimes\sigma_{z},
\end{align} where $h({\bf k})=\epsilon_{\bf k}\sigma_{0}+Ak_{x}\sigma_{x}+Ak_{y}\sigma_{y}+{\cal M}_{\bf k}\sigma_{z}$, $h^{*}(-{\bf k})=\epsilon_{\bf k}\sigma_{0}-Ak_{x}\sigma_{x}-Ak_{y}\sigma_{y}^{*}+{\cal M}_{\bf k}\sigma_{z}$, $\epsilon_{\bf k}=C-Dk^{2}$, ${\cal M}_{\bf k}=M-Bk^{2}$, $k^{2}=k_{x}^{2}+k_{y}^{2}$, $A$, $B$, $C$, $D$, and $M$ are the material specific parameters,  ${\bf k}=(k_x, k_y)$, $k_x$ and $k_y$ are the momenta of two-dimensional electron gas, $\sigma_{x,y,z}$ ($\tau_{x,y,z}$) are the Pauli matrices, and $\sigma_{0}$ ($\tau_{0}$) is the $2\times2$ identity matrix.
The parameters are adopted as~\cite{konig2008quantum} $A=364.5$ meV$\cdot$nm, $B=-686$ meV$\cdot$nm$^2$, $D=-512$ meV$\cdot$nm$^2$, $M=-10$ meV. Without loss of generality, we assume that $C=0$. 

In particular, the $h({\bf k})$ and $h^{*}(-{\bf k})$ in the matrix diagonal blocks of Eq.~(\ref{eq:H_BHZ}) provide the simplest ansatz that ensures time inversion symmetry in the model Hamiltonian.
In addition, $h({\bf k})$ and $h^{*}(-{\bf k})$ are equivalent to the two-dimensional Dirac model or Qi-Wu-Zhang (QWZ) model~\cite{qi2006topological,qi2011topological,chen2019finite}, which has a quantized Hall conductance without Landau levels. Thus, $h({\bf k})$ and $h^{*}(-{\bf k})$ correspond to two equal but opposite QAH Hamiltonians. 
Therefore, the net Hall conductance of the Hamiltonian (\ref{eq:H_BHZ}) is zero, but the spin Hall conductance for $h({\bf k})$ or $h^{*}(-{\bf k})$ is nonzero. This phenomenon corresponds to the QSH state. 
The BHZ model can be viewed as two copies of QWZ models with opposite Hall conductance.

\section{Floquet Hamiltonian}\label{3}

To realize QAH states in the Hamiltonian (\ref{eq:H_BHZ}), circularly polarized light can be used to break the time-reversal symmetry between the two spin blocks of the matrix (\ref{eq:H_BHZ}) to result in nonzero net Hall conductance.

The optical driving field can be expressed as ${\bf E}(t) = \partial{\bf A}(t)/\partial t = E( \cos(\omega t),  \cos(\omega t + \varphi), 0 )$, where ${\bf A}(t)={\bf A}(t+T)$ is the time periodic vector potential with period $T=2\pi/\omega$, $\omega$ is the frequency of the optical field, and $E$ is the amplitude of the optical field. The optical field is circularly polarized with $\varphi=\pm\pi/2$ and ${\bf A}(t)=A_{a}( \sin(\omega t),  \pm\cos(\omega t), 0 )$, where $A_a=E/\omega$ and $\pm$ corresponds to right- and left-handed circularly polarized waves, respectively. 
To guarantee the off-resonant regime in which the central Floquet band is far away from other replicas, the driven
frequency can be set as $\hbar\omega=150$ meV, which is much larger than the bandwidth~\cite{trevisan2022bicircular,qin2022phase,dabiri2021light,dabiri2021engineering,pervishko2018impact}. 

After irradiation with light, the photon-dressed effective Hamiltonian is given by
\begin{align}\label{eq:Ht}
{\cal H}({\bf k},t)= {\cal H}\left({\bf k} - \frac{e}{\hbar}{\bf A}(t)\right).
\end{align} 
Within the Floquet theory~\cite{oka2009photovoltaic,calvo2015floquet,chen2018floquet1,chen2018floquet2,du2022weyl,ning2022photoinduced,qin2022phase,wang2022floquet,jangjan2020floquet,jangjan2021topological} and $A_{0}=eA_{a}/\hbar$ which denotes the light intensity, the effective static Hamiltonian in the high-frequency regime can be expanded as
\begin{align}\label{eq:HF0}
{\cal H}^{(F)} =  {\cal H}_{0,0} + \frac{[{\cal H}_{0,-1}, {\cal H}_{0,1}]}{\hbar\omega} ,
\end{align} where we use ${\cal H}_{n,m} = \frac{1}{T} \int_{0}^{T}{\cal H}({\bf k},t) e^{i(n-m)\omega t}dt$ with $n$ and $m$ as integers, and the concrete analytical expressions for ${\cal H}_{0,0}$, ${\cal H}_{0,-1}$, and ${\cal H}_{0,1}$ can be found in Appendix \ref{A}.

From Eq.~(\ref{eq:HF0}), the Floquet Hamiltonian is given by
\begin{align}\label{eq:H_F}
{\cal H}^{(F)}({\bf k}) &\!=\!
\begin{pmatrix}
    h_{\uparrow}({\bf k}) &  0 \\
    0 & h_{\downarrow}({\bf k})
\end{pmatrix},
\end{align} where
$h_{\uparrow}({\bf k})
\!=\! -D(A_{0}^{2}+k^{2})\sigma_0\!+\!\left(\!\tilde{M}_{\uparrow}\!-\!Bk^{2} \!\right)\!\sigma_z 
\!+\!\tilde{A}_{\uparrow}(k_{x}\sigma_{x} \!+\! k_{y}\sigma_{y})$,
$h_{\downarrow}({\bf k}) 
\!=\! -D(A_{0}^{2}+k^{2})\sigma_0\!+\!\left(\!\tilde{M}_{\downarrow}\!-\!Bk^{2} \!\right)\!\sigma_z 
\!-\!\tilde{A}_{\downarrow}(k_{x}\sigma_{x} \!-\! k_{y}\sigma_{y})$, 
$\tilde{A}_{\uparrow}\!=\!\left[\!1\!-\! 2A_{0}^{2}B\sin\varphi/(\hbar\omega)\!\right]A$,
$\tilde{A}_{\downarrow}\!=\!\left[\!1\!+\! 2A_{0}^{2}B\sin\varphi/(\hbar\omega)\!\right]A$,
$\tilde{M}_{\uparrow}\!=\!M\!-\!BA_{0}^{2}\!-\! A_{0}^{2}A^{2}\sin\varphi/(\hbar\omega)$, and 
$\tilde{M}_{\downarrow}\!=\!M\!-\!BA_{0}^{2}\!+\! A_{0}^{2}A^{2}\sin\varphi/(\hbar\omega)$.
Note that it should not be taken for granted that optical driving will induce the QAH phase: when $\varphi=0$, we have $\tilde{M}_{\uparrow/\downarrow}\!=\!M\!-\!BA_{0}^{2}$ and $\tilde{A}_{\uparrow/\downarrow}=A$, so that the Floquet Hamiltonian (\ref{eq:H_F}) still satisfies the time-reversal symmetry, i.e., ${\cal T}{\cal H}^{(F)}({\bf k}){\cal T}^{-1}\!=\!{\cal H}^{(F)}(-{\bf k})$ with the time-reversal symmetry operator ${\cal T}\!=\!i\tau_{y}\otimes\sigma_{0}{\cal K} $~\cite{michetti2012helical}, where ${\cal K}$ is the operator of complex conjugation. 
However, when $\varphi\neq 0$, the terms containing $\varphi$ in $\tilde{M}_{\uparrow/\downarrow}$ and $\tilde{A}_{\uparrow/\downarrow}$ lead to ${\cal T}{\cal H}^{(F)}({\bf k}){\cal T}^{-1}\!\neq\!{\cal H}^{(F)}(-{\bf k})$ which breaks time-reversal symmetry of the Hamiltonian (\ref{eq:H_F}). The corresponding detailed derivations can be found in Appendix \ref{B}. This is the so-called time-reversal-symmetry-broken QSH insulator~\cite{yang2011time,chen2017disorder}.

The energy dispersions of the Floquet Hamiltonian (\ref{eq:H_F}) are given by
\begin{align}
E_{\uparrow,\pm}(k)&\!=\! -D(A_{0}^{2}+k^{2})
\!\pm\!\sqrt{(\tilde{M}_{\uparrow}\!-\!Bk^{2} )^{2} + \tilde{A}_{\uparrow}^{2}k^{2}}, \\
E_{\downarrow,\pm}(k)&\!=\! -D(A_{0}^{2}+k^{2})
\!\pm\!\sqrt{(\tilde{M}_{\downarrow}\!-\!Bk^{2} )^{2} + \tilde{A}_{\downarrow}^{2}k^{2}}.
\end{align}

To estimate the validity of the high-frequency expansions developed here quantitatively, we evaluate the maximum instantaneous energy of the time-dependent Hamiltonian (\ref{eq:Ht}) averaged over a period of the field $\frac{1}{T}\int_{0}^{T}dt~\text{max}\left\{\big|\big|H({\bf k},t)\big|\big|\right\}<\hbar\omega$ at the $\Gamma$ point ($k_x=k_y= 0$). The optical field parameters have to satisfy the condition $A_{0}A/(\hbar\omega)<1$. In the high-frequency regime $\omega \sim 2.2789\times10^{2}$ THz ($\hbar\omega=150$ meV), one can obtain $A_{0}\ll0.41$ nm$^{-1}$.
In the following numerical calculations, we have regularized the continuous model Hamiltonian (\ref{eq:H_F}) into the corresponding lattice Hamiltonian where topological invariants are well-defined (the detailed derivations can be found in Appendix \ref{C}). All the numerical results are based on the lattice regularized Hamiltonian, even though they all pertain to the region around the $\Gamma$ point.

\section{Energy dispersions}\label{4}

\begin{figure}[htpb]
\centering
\includegraphics[width=0.48\textwidth]{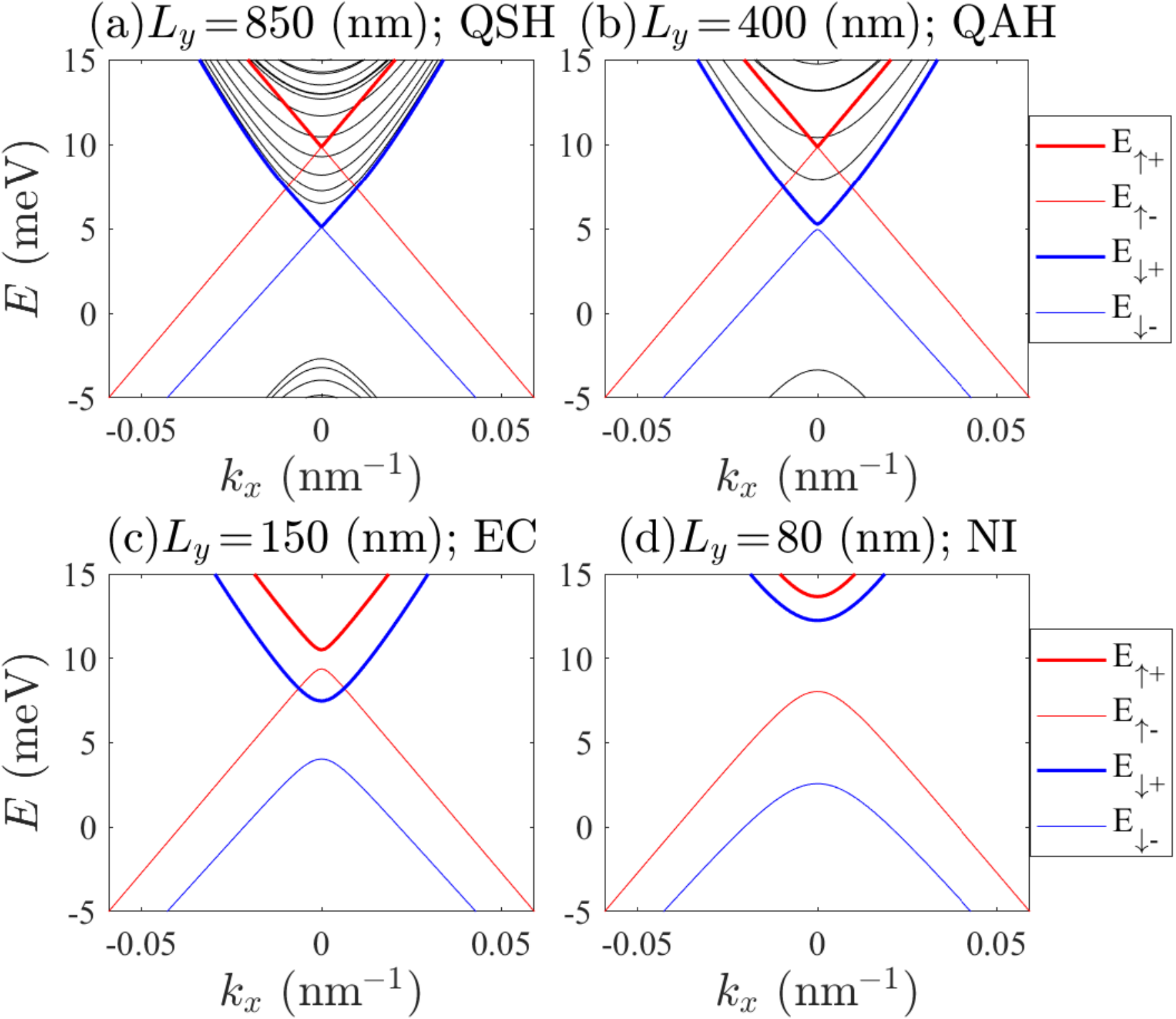}
\caption{(Color online) Energy bands for the Floquet Hamiltonian of the HgTe/(Hg,Cd)Te with different size $L_y$ under open boundary condition along the $y$ direction and periodic boundary conditions along the $x$ direction. (a) $L_y=850$ nm corresponds to quantum spin Hall (QSH). (b) $L_y=400$ nm corresponds to quantum anomalous Hall (QAH). (c) $L_y=150$ nm corresponds to edge conducting (EC). (d) $L_y=80$ nm corresponds to normal insulator (NI). Here, the definitions of $E_{\uparrow\pm}$ and $E_{\downarrow\pm}$ are the energies of two spin-up and spin-down edge bands under open boundary condition along the $y$ direction.
Here, QSH is topological protected for both spin-up and spin-down edge states, QAH is topological protected for one of them only, and EC for none of them, although they still exist unless broken by disorder.
The threshold of the energy gap for calculating the phase boundaries is 0.01 meV.
The other parameters are given as $A_{0}=0.06$ nm$^{-1}$, $a_x=a_y=1$ nm, $\varphi=\pi/2$, and $\hbar\omega=150$ meV. Here, $a_x$ and $a_y$ are the lattice constants along the $x$ and $y$ directions, respectively. 
Notice that we have mapped the continuous model Hamiltonian (\ref{eq:H_F}) into the lattice Hamiltonian (the detailed derivations can be found in Appendix \ref{C}).}  \label{Fig:E_tb}
\end{figure}

\begin{figure}[htpb]
\centering
\includegraphics[width=0.48\textwidth]{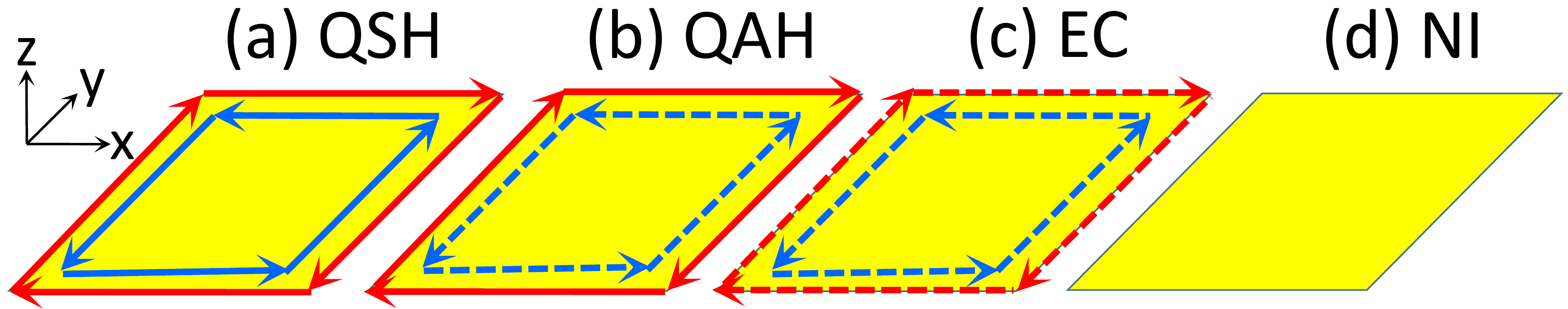}
\caption{(Color online) Schematic diagram (or physical picture) for the four phases with the clockwise (red lines) and anticlockwise (blue lines) charge currents of the spin-up and spin-down edge states, respectively.
(a) QSH state, (b) QAH state, (c) EC state, and (d) NI state.
The solid red line denotes the current for the spin-up edge states, the solid blue line denotes the current for the spin-down edge states, and the dashed red and blue lines denote the currents for the spin-up and spin-down edge states possible with the disorder, respectively. It means that the dashed lines refer to states that are not protected against disorder.}  \label{Fig:Schematic_2D}
\end{figure}

In order to explore how sufficiently small system sizes can lead to Floquet phase transitions in a QSH system, we calculate the tight-binding energy spectra of a HgTe/(Hg,Cd)Te quantum well with fixed light intensity in a stripe geometry with sample width $L_y$ along the $y$ direction, and the system length in the $x$ direction is infinite. 
The energy gap opening and closing in the energy dispersions of the edge states is the most important feature of finite-size effects with fixed light intensity.

As shown in Fig.~\ref{Fig:E_tb}, we calculate the energy spectra of the HgTe/(Hg,Cd)Te with fixed light intensity $A_{0}=0.06$ nm$^{-1}$ and different width $L_{y}$ under the open boundary condition along the $y$ direction and periodic boundary conditions along the $x$ direction (the detailed derivations can be found in Appendix \ref{C}). In addition, we give the detailed derivations of the tight-binding model under the open boundary condition along the $x$ direction and periodic boundary conditions along the $y$ direction in Appendix \ref{D}.
In Fig.~\ref{Fig:E_tb}, the red thick line indicates the upper band $E_{\uparrow+}$ of the spin-up edge states, the red thin line indicates the lower band $E_{\uparrow-}$ of the spin-up edge states, the blue thick line indicates the upper band $E_{\downarrow+}$ of the spin-down edge states, and the blue thin line indicates the lower band $E_{\downarrow-}$ of the spin-down edge states.

To illustrate the physical picture for the four phases with different system widths in Fig.~\ref{Fig:E_tb}, we provide a schematic diagram for the four phases with the clockwise (red lines) and anticlockwise (blue lines) charge currents of the spin-up and spin-down edge states, respectively, as shown in Fig.~\ref{Fig:Schematic_2D}.

In Fig.~\ref{Fig:E_tb}(a), the system width is $L_y=850$ nm which is sufficiently large that there is no coupling between edge states. It is found that the energy dispersions for both spin-up and spin-down edge states are gapless and each spin state provides a charge conductance $e^2/h$, as shown in Fig.~\ref{Fig:Schematic_2D}(a).
This indicates that the system is in the QSH state which is time-reversal symmetry broken due to circularly polarized light; the corresponding charge conductance along the $x$ direction is $2e^2/h$, as shown in Fig.~\ref{Fig:Schematic_2D}(a), with the contributions from both spin-up and spin-down charge currents flowing from higher potential to lower potential; and it is robust against the disorder.

In Fig.~\ref{Fig:E_tb}(b), the system width is decreased to $L_y=400$ nm, where the coupling between the spin-down edge states is strong enough to open an energy gap in the corresponding energy dispersions, while the energy dispersions of the spin-up edge states remain gapless which provides a charge conductance $e^2/h$.
As a result, the spin-down edge states can easily be broken by the disorder.
This is because the transport electrons in the spin-down edge channels can possibly be backscattered by impurities. Hence, the charge conductance along the $x$ direction in this state ranges from $e^2/h$ to $2e^2/h$ with disorder, as shown in Fig.~\ref{Fig:Schematic_2D}(b). 
This state is called the QAH state~\cite{fu2014finite}.

In Fig.~\ref{Fig:E_tb}(c), when the system width decreases to $L_y=150$ nm, all the energy dispersions of the edge states become gapped. 
Particularly, the lower band of the spin-up edge states is higher than the upper band of the spin-down edge states near the $k_x=0$ point. Therefore, there are always conducting edge channels near the Fermi level which is in the bulk gap. This state is called the EC state~\cite{fu2014finite}, in which the charge currents through the edge can be killed by disorder. So the corresponding charge conductance along the $x$ direction ranges from $0$ to $2e^2/h$, as shown in Fig.~\ref{Fig:Schematic_2D}(c).
In Fig.~\ref{Fig:E_tb}(d), by further decreasing the system width to $L_y=80$ nm, there is a bulk energy gap in the energy dispersions and we get a NI state which corresponds to Fig.~\ref{Fig:Schematic_2D}(d) and is induced by finite-size effects.

\begin{figure}[htpb]
\centering
\includegraphics[width=0.48\textwidth]{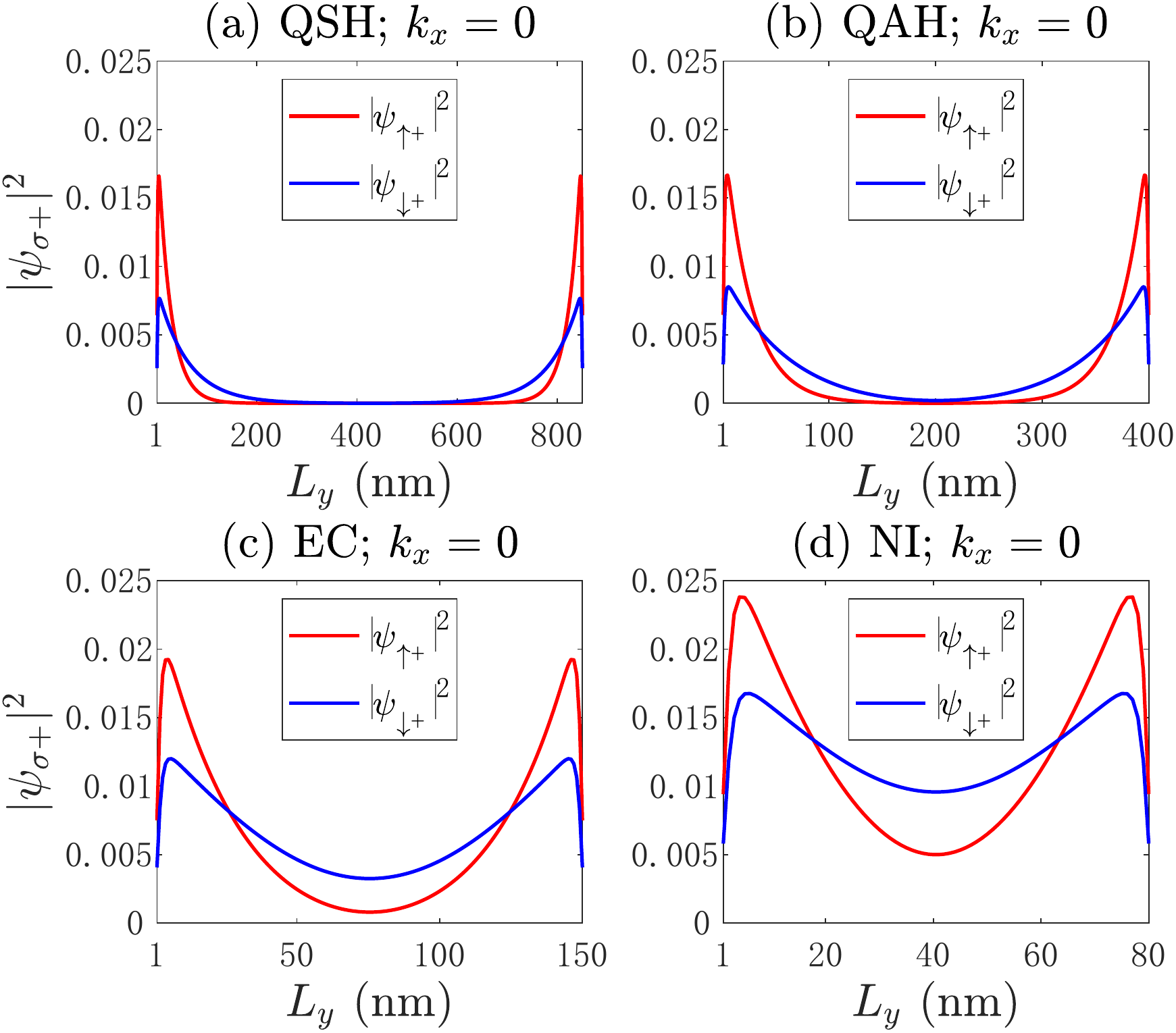}
\caption{(Color online) Probability distributions of the edge states for the Floquet Hamiltonian of the HgTe/(Hg,Cd)Te with different size $L_y$ under $k_{x}=0$. The other parameters are the same as those in Fig.~\ref{Fig:E_tb}.
Notice that the subscript $+$ denotes the edge states of $E_{\sigma+}$ with $\sigma=\uparrow,\downarrow$.
The red solid curve denotes the probability distribution for the spin-$\uparrow$ edge state and the blue solid curve denotes the one for the spin-$\downarrow$ edge state. The density distributions of the edge states for $E_{\sigma-}$ can be found in Appendix \ref{E}.}  \label{Fig:Vz_tb}
\end{figure}

Meanwhile, along the $y$ direction, the edge states have a probability distribution in the sense of the wave function of edge states. As shown in Fig.~\ref{Fig:Vz_tb}, one can obtain a clear idea of how well-located edge states are forced to be mixed by changing $L_{y}$.
Here, the red solid curve denotes the probability distribution for the spin-$\uparrow$ edge state and the blue solid curve denotes the one for the spin-$\downarrow$ edge state.
It is indicated from Fig.~\ref{Fig:Vz_tb}(a) that the wave functions for both spin-$\uparrow$ and spin-$\downarrow$ edge states are well localized at the left and right edges of the system with large system width $L_y=850$ nm where there is no coupling between the edge states. When the system width $L_y$ becomes smaller and smaller, as shown in Figs.~\ref{Fig:Vz_tb}(b)--\ref{Fig:Vz_tb}(d), it is found that the edge localizations of the wave functions of the edge states get worse and worse.

\begin{figure}[htpb]
\centering
\includegraphics[width=0.48\textwidth]{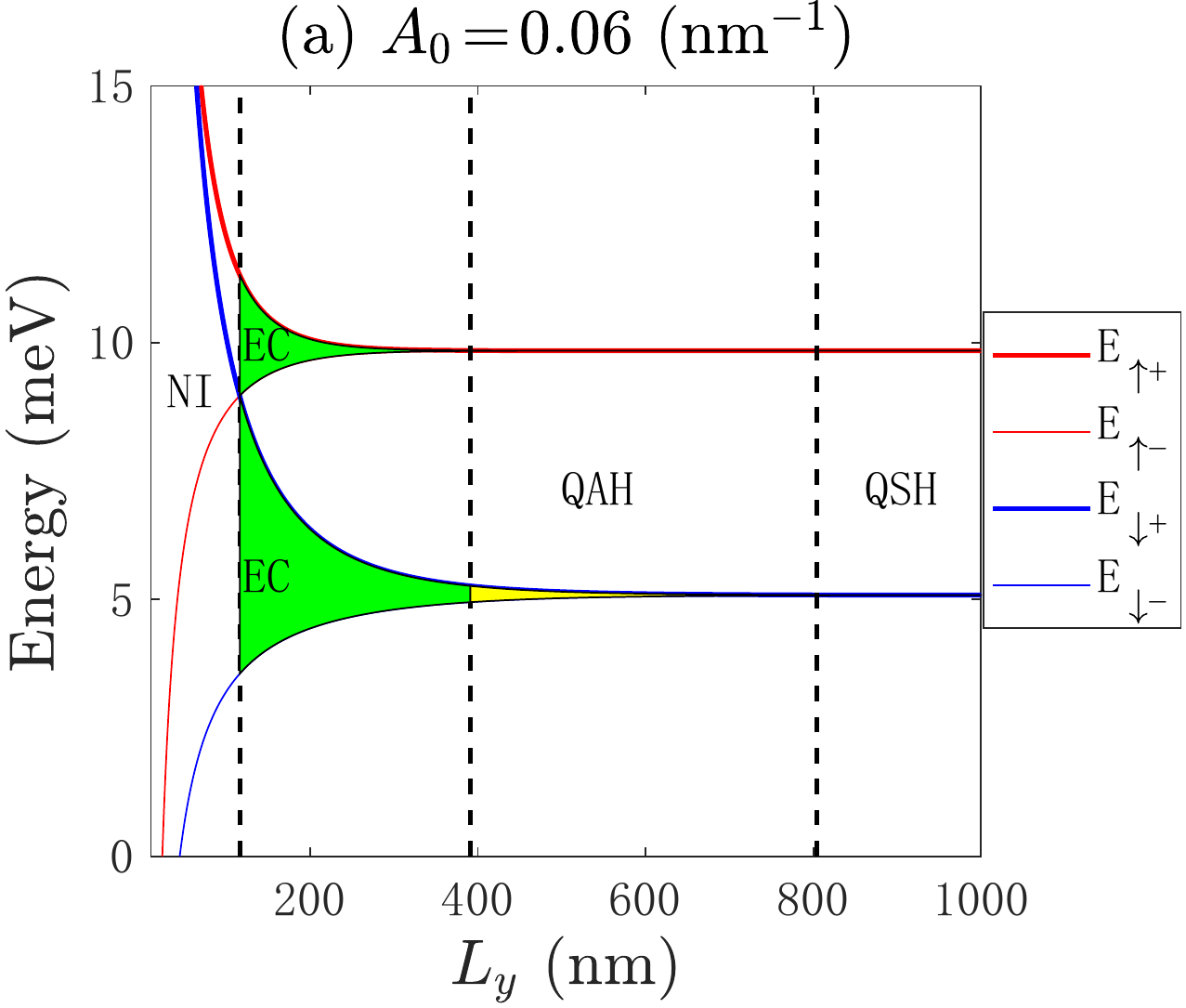}
\includegraphics[width=0.48\textwidth]{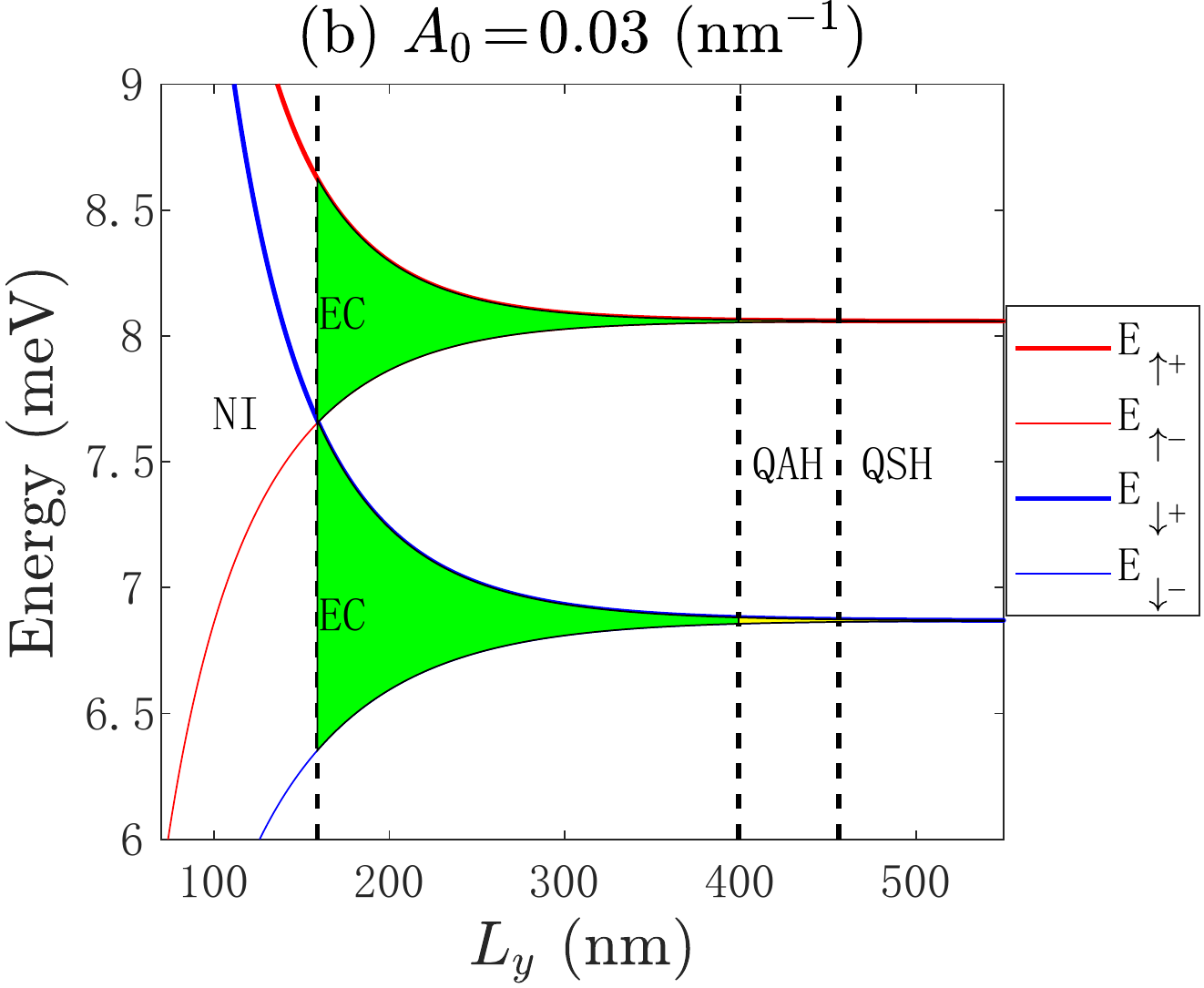}
\caption{(Color online) Band energies $E_{\uparrow\pm}$ and $E_{\downarrow\pm}$ at the zero momentum point $k_x=0$ as a function of the size $L_y$ with fixed (a) $A_{0}=0.06$ nm$^{-1}$ and (b) $A_{0}=0.03$ nm$^{-1}$ under the open boundary condition along the $y$ direction and periodic boundary conditions along the $x$ direction.
The threshold of the energy gap for calculating the phase boundaries is $0.01$ meV.
The other parameters are the same as those in Fig.~\ref{Fig:E_tb}. }  \label{Fig:Phase2D_E_Ly_yOBC}
\end{figure}

We furthermore investigate phase crossover due to adjusting the finite system width, with light intensity fixed.
In Fig.~\ref{Fig:Phase2D_E_Ly_yOBC}, the definitions of $E_{\uparrow\pm}$ and $E_{\downarrow\pm}$ are the energies of two spin-up and spin-down edge bands at the zero momentum point $k_x=0$ under the open boundary condition along the $y$ direction. Meanwhile, the red thick line indicates the upper band $E_{\uparrow+}$ of the spin-up edge states, the red thin line indicates the lower band $E_{\uparrow-}$ of the spin-up edge states, the blue thick line indicates the upper band $E_{\downarrow+}$ of the spin-down edge states, and the blue thin line indicates the lower band $E_{\downarrow-}$ of the spin-down edge states.

We calculate the energies ($E_{\uparrow\pm}$ and $E_{\downarrow\pm}$) of the spin-up and spin-down edge states at $k_x=0$ as a function of the system width $L_y$ as shown in Fig.~\ref{Fig:Phase2D_E_Ly_yOBC}(a) with the same parameters as those in Fig.~\ref{Fig:E_tb}. 
When $L_y$ is very small, the energy gap for both bulk and edge states is large and the system is insulating in the NI state. By gradually increasing the width $L_y$, once the lower energy band of the spin-up edge states touches the upper energy band of the spin-down edge states, the system enters into the EC state.
Particularly, we notice that there is a critical point which separates the EC state from the NI state, as shown in Fig.~\ref{Fig:Phase2D_E_Ly_yOBC}. In the EC state, the lower energy band of the spin-up edge states is higher than the upper energy band of the spin-down edge states, and the energy dispersions for both spin-up and spin-down edge states are gapped. 
By further increasing the width $L_y$, the energy gap of the spin-up edge states vanishes but the energy dispersion for the spin-down edge states is still gapped. At this time, the system enters into a QAH state. 
When $L_y$ is very large, the system enters into a QSH state, where the energy dispersions for both spin-up and spin-down edge states are gapless.

Different phases are determined by the edge states and energy gaps. Here, we consider the finite-size effect on the edge states and energy gaps. However, the Chern number and Bott index are used to describe the topological properties of an infinite-size system under periodic boundary conditions both along the $x$ and $y$ directions. Therefore, for a small-size system under open boundary condition along the $y$ direction here, both the Chern number and Bott index are inapplicable. Concretely, when the system size is very small, the coupling between the edge states becomes very strong and the edge states are destroyed by opening a gap, as shown in Fig.~\ref{Fig:Vz_tb}(d). In addition, Ref.~\cite{zhou2008finite} also mentioned that the finite-size effect can destroy the quantum spin Hall effect phase without a topological index transition. Next, we will reply to the applicability of our approach to the phase transition regime. For finite small sizes, our phase transitions belong to continuous phase transitions which are in general crossovers, and we can only use the energy gaps of the edge states to determine the phase transitions quantitatively. Meanwhile, as mentioned in Ref.~\cite{fu2014finite}, another applicable method is to calculate the transport or conductance of the finite-size system by using the Landauer-B\"uttiker formalism~\cite{landauer1970electrical,buttiker1988absence}. For example, Ref.~\cite{li2011size} has calculated the transport properties of a finite-size system in HgTe/CdTe quantum wells without optical pumping. We will investigate the transport properties of a finite-size Floquet system in the future.

\section{Overall phase diagram}\label{5}

In this section, we will investigate the overall phase diagram of the system.

\begin{figure}[htpb]
\centering
\includegraphics[width=0.48\textwidth]{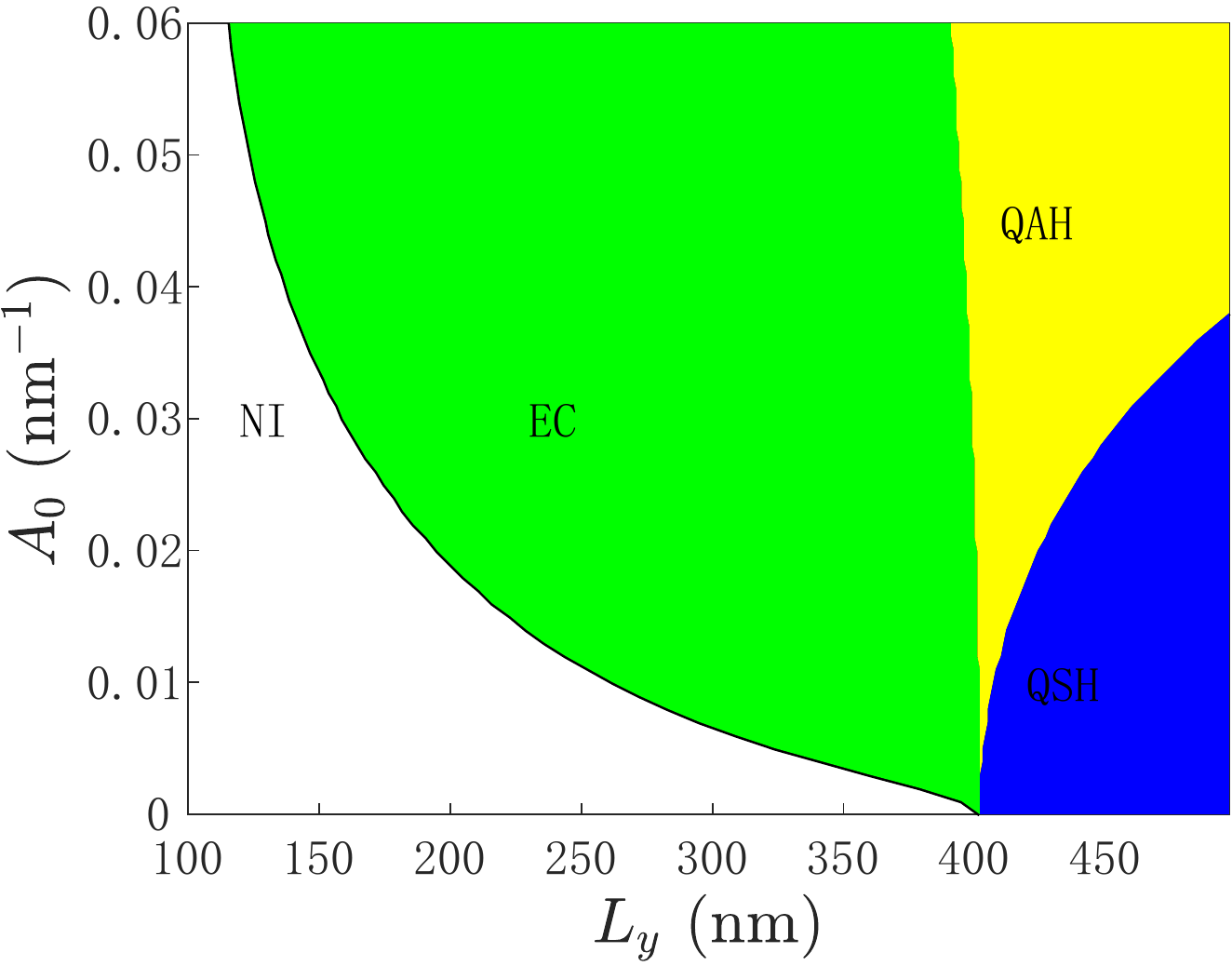}
\caption{(Color online) Phase diagram in the system width vs. light intensity ($L_y$, $A_0$) plane under the open boundary condition along the $y$ direction and periodic boundary conditions along the $x$ direction, as determined by the long-$x$-wavelength (small-$k_x$) limit.  
There are four kinds of phase regimes, i.e., NI, EC, QAH, and QSH are shown.
The white zone denotes the NI regime, the green zone denotes the EC regime, the yellow zone denotes the QAH regime, and the blue zone denotes the QAH regime.  
The threshold of the energy gap for calculating the phase boundaries is $0.01$ meV, which is also used
as a criterion of the energy gap opening.
The other parameters are the same as those in Fig.~\ref{Fig:E_tb}.}  \label{Fig:Phase3D_Delta_yOBC}
\end{figure}

Based on the energy gap for spin-up and spin-down edge states, we plot the whole phase diagram with two tunable parameters: light intensity $A_0$ and system width $L_y$ as shown in Fig.~\ref{Fig:Phase3D_Delta_yOBC}.
There are four kinds of phase regimes, i.e., NI, EC, QAH, and QSH, as shown in Fig.~\ref{Fig:Phase3D_Delta_yOBC}.
An important feature is that each transport regime has its own permitted width range with fixed light intensity. For example, the width for the EC state is in the range from $116$ to $391$ nm with fixed $A_0=0.06$ nm$^{-1}$. The permitted width range for different transport states is important because it can help us to find the desired transport regime by choosing an appropriate device width and intensity of light.

\section{Summary}\label{6}

Even though Floquet engineering had been extensively used to engineer various topological phases that are elusive in the static regime, our proposal instead engineers QAH and other phases in a realistic QSH setup that takes into account the hybridization effects due to the finite size of the realistic samples. We numerically investigate the light-induced topological phases and finite-size topological phase crossovers in a QSH system with high-frequency pumping optics, taking the HgTe quantum well as the paradigmatic realistic model. 
We find that the coupling of edge states depends on both the system width and the light intensity.
Furthermore, we can get four different topological transport regimes by choosing the proper width and light intensity. Since the intensity of light can easily be continuously tuned in experiments, our proposal that uses circularly polarized light instead of magnetic doping provides a feasible route towards realizing QAH states in experiments.

With state-of-the-art nanofabrication, it is possible to engineer realistic quantum materials to reduce its size in the length scale presented in our figures. For example, in Ref.~\cite{qiu2022mesoscopic}, it is reported that the quantum anomalous Hall (QAH) devices can be fabricated as miniaturized with channel widths down to 600 nm in six quintuple layers of Cr$_{0.12}$(Bi$_{0.26}$Sb$_{0.62}$)$_{2}$Te$_{3}$ film which were epitaxially deposited onto GaAs (111) substrates. Furthermore, they measured the longitudinal and transverse resistivities in four devices with a channel width of 0.6, 1.5, 3, and 5 $\mu$m~\cite{qiu2022mesoscopic}. Meanwhile, Ref.~\cite{zhou2022confinement} demonstrates that the QAH effect persists in a Hall bar device with width 72 nm in three quintuple layers of Cr-doped (Bi,Sb)$_{2}$Te$_{3}$, four quintuple layers of (Bi,Sb)$_{2}$Te$_{3}$, and three quintuple layers of Cr-doped (Bi,Sb)$_{2}$Te$_{3}$~\cite{zhao2020tuning,zhao2022zero}, and they measured the transverse resistance and longitudinal resistance in Hall bar devices with widths from 300 nm to 10 $\mu$m. Technologically, first, by using photolithography, the samples can be patterned into a Hall bar with width 10 $\mu$m~\cite{qiu2022mesoscopic,zhou2022confinement}. Then by using another experimental technology, i.e., electron-beam lithography, the realization of the QAH bar devices with widths from 10 $\mu$m down to 72 nm are further fabricated~\cite{zhou2022confinement}.

\begin{acknowledgments}
We acknowledge helpful discussions with Hao-Jie Lin and Xiao-Bin Qiang. 
This work is supported by the Singapore National Research Foundation (Grant No.~NRF2021-QEP2-02-P09). 
F.Q. acknowledges support from the National Natural Science Foundation of China (Grants No.~11404106), and the project funded by the China Postdoctoral Science Foundation (Grants No.~2019M662150 and No.~2020T130635) before he joined NUS.
R.C. acknowledges support from the project funded by the China Postdoctoral Science Foundation (Grant No.~2019M661678).
\end{acknowledgments}

\bibliography{references_floquet2D}

\clearpage
\begin{widetext}
\appendix

\section{Expressions for ${\cal H}_{0,0}$, ${\cal H}_{0,-1}$, and ${\cal H}_{0,1}$}\label{A}

The motivation for Appendix \ref{A} is to show the concrete analytical expressions for ${\cal H}_{0,0}$, ${\cal H}_{0,-1}$, and ${\cal H}_{0,1}$ in the Floquet Hamiltonian (\ref{eq:HF0}) of the main text.
The concrete analytical expressions for ${\cal H}_{0,0}$, ${\cal H}_{0,-1}$, and ${\cal H}_{0,1}$ in the Floquet Hamiltonian (\ref{eq:HF0}) are given as 
\begin{align}
{\cal H}_{0,0}
&\!=\! [C - D(A_{0}^{2}+k_{x}^{2}+k_{y}^{2})]\tau_{0}\otimes\sigma_{0} + Ak_{x}\tau_{z}\otimes\sigma_{x} + [M - B(A_{0}^{2}+k_{x}^{2}+k_{y}^{2})]\tau_{0}\otimes\sigma_{z} + Ak_{y}\tau_{0}\otimes\sigma_{y} \nonumber\\
&\!=\! {\cal H}({\bf k}) - DA_{0}^{2}\tau_{0}\otimes\sigma_{0} - BA_{0}^{2}\tau_{0}\otimes\sigma_{z},\label{eq:H00}\\
\!{\cal H}_{0,-1}
&\!=\! iDA_{0}(k_{x} \!+\! e^{-i\varphi}k_{y})\tau_{0}\otimes\sigma_{0} \!+\! iBA_{0}(k_{x} \!+\! e^{-i\varphi}k_{y})\tau_{0}\otimes\sigma_{z} \!-\!\frac{i}{2}AA_{0}\tau_{z}\otimes\sigma_{x} \!-\!\frac{i}{2}AA_{0}e^{-i\varphi}\tau_{0}\otimes\sigma_{y}, \label{eq:H0-1}\\
\!{\cal H}_{0,1}
&\!=\! \!-\!iDA_{0}(k_{x} \!+\! e^{i\varphi}k_{y})\tau_{0}\otimes\sigma_{0} 
\!-\! iBA_{0}(k_{x} \!+\! e^{i\varphi}k_{y})\tau_{0}\otimes\sigma_{z} 
\!+\!\frac{i}{2}AA_{0}\tau_{z}\otimes\sigma_{x} \!+\!\frac{i}{2}AA_{0}e^{i\varphi}\tau_{0}\otimes\sigma_{y}.\label{eq:H01}
\end{align}

\section{Time-reversal symmetry}\label{B}

The motivation for Appendix \ref{B} is to prove that when $\varphi=0$, the Floquet Hamiltonian (\ref{eq:H_F}) in the main text still satisfies the time-reversal symmetry. However, when $\varphi \neq 0$, the time-reversal symmetry of the Hamiltonian (\ref{eq:H_F}) is broken.

The Floquet Hamiltonian~(\ref{eq:H_F}) under time-reversal transformation becomes
\begin{align}\label{eq:time_reversal}
&~~{\cal T}{\cal H}^{(F)}({\bf k}){\cal T}^{-1} \nonumber\\
&\!=\!i\tau_{y}\otimes\sigma_{0}{\cal K}\left[ -D(A_{0}^{2}+k^{2})\tau_{0}\otimes\sigma_{0} + A(k_{x}\tau_{z}\otimes\sigma_{x} + k_{y}\tau_{0}\otimes\sigma_{y}) + (\!M\!-\!BA_{0}^{2}\!-\!Bk^{2})\tau_{0}\otimes\sigma_{z}  \right]{\cal K}^{-1}(i\tau_{y}\otimes\sigma_{0})^{-1} \nonumber\\
&\!-\!\frac{A_{0}^{2}\sin\varphi}{\hbar\omega}i\tau_{y}\otimes\sigma_{0}{\cal K}\left[ 2AB(k_{x}\tau_{0}\otimes\sigma_{x} + k_{y}\tau_{z}\otimes\sigma_{y}) + A^{2}\tau_{z}\otimes\sigma_{z} \right]{\cal K}^{-1}(i\tau_{y}\otimes\sigma_{0})^{-1} \nonumber\\
&\!=\! -D(A_{0}^{2}+k^{2})\tau_{0}\otimes\sigma_{0} - A(k_{x}\tau_{z}\otimes\sigma_{x} + k_{y}\tau_{0}\otimes\sigma_{y}) + (\!M\!-\!BA_{0}^{2}\!-\!Bk^{2})\tau_{0}\otimes\sigma_{z}  \nonumber\\
&\!-\!\frac{A_{0}^{2}\sin\varphi}{\hbar\omega}\left[ 2AB(k_{x}\tau_{0}\otimes\sigma_{x} + k_{y}\tau_{z}\otimes\sigma_{y}) - A^{2}\tau_{z}\otimes\sigma_{z} \right] \nonumber\\
&\!=\! {\cal H}^{(F)}(-{\bf k}) \!-\!\frac{2A_{0}^{2}\sin\varphi}{\hbar\omega}\left[ 2AB(k_{x}\tau_{0}\otimes\sigma_{x} + k_{y}\tau_{z}\otimes\sigma_{y}) - A^{2}\tau_{z}\otimes\sigma_{z} \right] ,
\end{align}
where ${\cal T}=i\tau_{y}\otimes\sigma_{0}{\cal K}$~\cite{michetti2012helical} is the time-reversal operator with the complex conjugate operator ${\cal K}$, we have ${\cal K}{\cal H}^{(F)}({\bf k}){\cal K}^{-1}={\cal H}^{(F)*}({\bf k})$, and we use
\begin{align}
&(i\tau_{y}\otimes\sigma_{0})(\tau_{0}\otimes\sigma_{0})(i\tau_{y}\otimes\sigma_{0})^{-1}=\tau_{0}\otimes\sigma_{0},\\
&(i\tau_{y}\otimes\sigma_{0})(\tau_{z}\otimes\sigma_{x})(i\tau_{y}\otimes\sigma_{0})^{-1}=-\tau_{z}\otimes\sigma_{x} ,\\
&(i\tau_{y}\otimes\sigma_{0})(K\tau_{0}\otimes\sigma_{y}K^{-1})(i\tau_{y}\otimes\sigma_{0})^{-1}=-\tau_{0}\otimes\sigma_{y} ,\\
&(i\tau_{y}\otimes\sigma_{0})(\tau_{0}\otimes\sigma_{z} )(i\tau_{y}\otimes\sigma_{0})^{-1}= \tau_{0}\otimes\sigma_{z},\\
&(i\tau_{y}\otimes\sigma_{0})(\tau_{0}\otimes\sigma_{x} )(i\tau_{y}\otimes\sigma_{0})^{-1}= \tau_{0}\otimes\sigma_{x},\\
&(i\tau_{y}\otimes\sigma_{0})(K\tau_{z}\otimes\sigma_{y}K^{-1})(i\tau_{y}\otimes\sigma_{0})^{-1}= \tau_{z}\otimes\sigma_{y}, \\
&(i\tau_{y}\otimes\sigma_{0})(\tau_{z}\otimes\sigma_{z} )(i\tau_{y}\otimes\sigma_{0})^{-1}= -\tau_{z}\otimes\sigma_{z}.
\end{align}
As a result of Eq.~(\ref{eq:time_reversal}), if $\varphi=0$, we have ${\cal T}{\cal H}^{(F)}({\bf k}){\cal T}^{-1}={\cal H}^{(F)}(-{\bf k})$ which shows a time-reversal symmetry.
However, for $\varphi\neq0$, Eq.~(\ref{eq:time_reversal}) shows that the time-reversal symmetry is broken.

\section{Tight-binding model under the open boundary condition along the $y$ direction and periodic boundary conditions along the $x$ direction}\label{C}

The motivation of Appendix \ref{C} is to derive the analytical expression of the tight-binding model Hamiltonian under the open boundary condition along the $y$ direction and periodic boundary conditions along the $x$ direction for the further numerical calculations in the main text.

In a lattice, one makes the following replacements~\cite{shen2012topological}:
\begin{align}
&k_{j}\rightarrow\frac{1}{a_j}\sin(k_{j}a_{j}),\\
&k_{j}^{2}\rightarrow\frac{2}{a_j^2}[1-\cos(k_{j}a_{j})],
\end{align} where $j=x,y,z$, $a_j$ is the lattice constant along the $j$ direction, $\sin(k_{j}a_{j})=\frac{e^{ik_{j}a_{j}}-e^{-ik_{j}a_{j}}}{2i}$, and $\cos(k_{j}a_{j})=\frac{e^{ik_{j}a_{j}}+e^{-ik_{j}a_{j}}}{2}$,
\begin{align}
k^{2}\rightarrow\frac{2}{a_x^2}[1-\cos(k_{x}a_{x})] + \frac{2}{a_y^2}[1-\cos(k_{y}a_{y})],
\end{align} where $k^{2}=k_{x}^{2}+k_{y}^{2}$.
In this way, the hopping terms in the lattice model only exist between the nearest-neighbor sites.
With this mapping, one obtains the following tight-binding model with high-frequency pumping in the basis $(c_{{\bf k},+,\uparrow}, c_{{\bf k},-,\uparrow}, c_{{\bf k},+,\downarrow}, c_{{\bf k},-,\downarrow})^{T}$ as 
\begin{align}\label{eq:HFfilm_supp}
\!{\cal H}^{(F)}({\bf k})\!=\! \left(
  \begin{array}{cccc}
    h_{\uparrow}({\bf k}) & 0  \\
    0 & h_{\downarrow}({\bf k})
  \end{array}
\right),
\end{align}
where
\begin{align}
h_{\uparrow}({\bf k})
&\!=\! \!-\! D\left\{A_{0}^{2} \!+\! \frac{2}{a_x^2}[1\!-\!\cos(k_{x}a_{x})] \!+\! \frac{2}{a_y^2}[1\!-\!\cos(k_{y}a_{y})] \right\} \sigma_0 \nonumber\\
&\!+\! \left\{M \!-\! B\left[A_{0}^{2} \!+\! \frac{2}{a_x^2}[1\!-\!\cos(k_{x}a_{x})] \!+\! \frac{2}{a_y^2}[1\!-\!\cos(k_{y}a_{y})] \right] 
\!-\! \frac{A_{0}^{2}A^{2}\sin\varphi}{\hbar\omega} \right\}\sigma_z \nonumber\\
&\!+\! \left(\!1\!-\! \frac{2A_{0}^{2}B\sin\varphi}{\hbar\omega}\!\right)A\left[\frac{1}{a_x}\sin(k_{x}a_{x}) \!-\! \frac{i}{a_y}\sin(k_{y}a_{y}) \right]\sigma_{+} \nonumber\\
&\!+\! \left(\!1\!-\! \frac{2A_{0}^{2}B\sin\varphi}{\hbar\omega}\!\right)A \left[\frac{1}{a_x}\sin(k_{x}a_{x}) \!+\! \frac{i}{a_y}\sin(k_{y}a_{y}) \right]\sigma_{-}, 
\end{align}
\begin{align}
h_{\downarrow}({\bf k})
&\!=\! \!-\! D\left\{A_{0}^{2} \!+\! \frac{2}{a_x^2}[1\!-\!\cos(k_{x}a_{x})] \!+\! \frac{2}{a_y^2}[1\!-\!\cos(k_{y}a_{y})] \right\} \sigma_0  \nonumber\\
&\!+\! \left\{M \!-\! B\left[A_{0}^{2} \!+\! \frac{2}{a_x^2}[1\!-\!\cos(k_{x}a_{x})] \!+\! \frac{2}{a_y^2}[1\!-\!\cos(k_{y}a_{y})] \right] \!+\! \frac{A_{0}^{2}A^{2}\sin\varphi}{\hbar\omega} \right\}\sigma_z \nonumber\\
&\!-\! \left(\!1\!+\! \frac{2A_{0}^{2}B\sin\varphi}{\hbar\omega}\!\right)A\left[\frac{1}{a_x}\sin(k_{x}a_{x}) \!+\! \frac{i}{a_y}\sin(k_{y}a_{y}) \right]\sigma_{+}\nonumber\\
&\!-\! \left(\!1\!+\! \frac{2A_{0}^{2}B\sin\varphi}{\hbar\omega}\!\right)A \left[\frac{1}{a_x}\sin(k_{x}a_{x}) \!-\! \frac{i}{a_y}\sin(k_{y}a_{y}) \right]\sigma_{-}.
\end{align}

Performing the Fourier transformation, one obtains
\begin{align}
c_{{\bf k},s,\sigma}&=\frac{1}{\sqrt{N_{y}}}\sum_{k_x,j_y}^{N_y}e^{-ik_{y}j_{y}a_{y}}c_{k_x,j_y,s,\sigma},\\
c_{{\bf k},s,\sigma}^{\dagger}&=\frac{1}{\sqrt{N_{y}}}\sum_{k_x,j_y}^{N_y}e^{ik_{y}j_{y}a_{y}}c_{k_x,j_y,s,\sigma}^{\dagger}.
\end{align}
\begin{align}
C_{{\bf k},\sigma}&=\begin{pmatrix}
c_{{\bf k},+,\sigma} \\
c_{{\bf k},-,\sigma}
\end{pmatrix}=\frac{1}{\sqrt{N_{y}}}\sum_{k_x,j_y}^{N_y}e^{-ik_{y}j_{y}a_{y}}C_{k_x,j_y,\sigma},\\
C_{{\bf k},\sigma}^{\dagger}&=\begin{pmatrix}
c_{{\bf k},+,\sigma}^{\dagger} &
c_{{\bf k},-,\sigma}^{\dagger}
\end{pmatrix}=\frac{1}{\sqrt{N_{y}}}\sum_{k_x,j_y}^{N_y}e^{ik_{y}j_{y}a_{y}}C_{k_x,j_y,\sigma}^{\dagger}.
\end{align} 

\begin{align}
&~~\sum_{{\bf k},s}C_{{\bf k},\uparrow}^{\dagger}\left[h_{\uparrow}({\bf k})\right]C_{{\bf k},\uparrow}\nonumber\\ 
&\!=\!\!-\!D\sum_{k_x,j_y}\!\left[\! A_{0}^{2} \!+\! \frac{2}{a_x^2}\!+\!\frac{2}{a_y^2}\!-\!\frac{2}{a_x^2}\cos(k_xa_x)\!\right]\!C_{k_x,j_y,\uparrow}^{\dagger}\sigma_{0}C_{k_x,j_y,\uparrow} 
\!+\! \frac{D}{a_y^2}\sum_{k_x,j_y}\!\left[\!C_{k_x,j_y,\uparrow}^{\dagger}\sigma_{0}C_{k_x,j_y+1,\uparrow}\!+\! {\rm h.c.}\!\right]\! \nonumber\\
&\!+\!\sum_{k_x,j_y}\!\left[\!M\!-\!BA_{0}^{2}\!-\!\frac{2B}{a_x^2}\!-\!\frac{2B}{a_y^2}\!+\!\frac{2B}{a_x^2}\cos(k_xa_x) \!-\! \frac{A_{0}^{2}A^{2}\sin\varphi}{\hbar\omega}\!\right]\!C_{k_x,j_y,\uparrow}^{\dagger}\sigma_{z}C_{k_x,j_y,\uparrow}\!+\! \frac{B}{a_y^2}\sum_{k_x,j_y}\!\left[\!C_{k_x,j_y,\uparrow}^{\dagger}\!\sigma_{z}\!C_{k_x,j_y+1,\uparrow}\!+\! {\rm h.c.}\!\right]\! \nonumber\\
&\!+\!\left(\!1\!-\! \frac{2A_{0}^{2}B\sin\varphi}{\hbar\omega}\!\right)\frac{A}{a_x}\sum_{k_x,j_y}\sin(k_xa_x) \! \left[\!C_{k_x,j_y,\uparrow}^{\dagger}(\sigma_{+}+\sigma_{-})C_{k_x,j_y,\uparrow} \!\right]\! \nonumber\\
&\!+\!\left(\!1\!-\! \frac{2A_{0}^{2}B\sin\varphi}{\hbar\omega}\!\right) \frac{A}{2a_y} 
\sum_{k_x,j_y}\!\left[\!C_{k_x,j_y,\uparrow}^{\dagger}(\sigma_{-}-\sigma_{+})C_{k_x,j_y+1,\uparrow} 
\!+\! C_{k_x,j_y+1,\uparrow}^{\dagger}(\sigma_{+}-\sigma_{-})C_{k_x,j_y,\uparrow}\!\right],
\end{align} where 
\begin{align}
&C_{k_x,j_y,\uparrow}^{\dagger}(\sigma_{-}-\sigma_{+})C_{k_x,j_y+1,\uparrow}\nonumber\\
&=
\begin{pmatrix}
c_{k_x,+,j_y,\uparrow}^{\dagger} &
c_{k_x,-,j_y,\uparrow}^{\dagger}
\end{pmatrix}
\begin{pmatrix}
0 & -1 \\
1 & 0
\end{pmatrix}
\begin{pmatrix}
c_{k_x,+,j_y+1,\uparrow} \\
c_{k_x,-,j_y+1,\uparrow}
\end{pmatrix} =\begin{pmatrix}
c_{k_x,-,j_y,\uparrow}^{\dagger} &
-c_{k_x,+,j_y,\uparrow}^{\dagger}
\end{pmatrix}
\begin{pmatrix}
c_{k_x,+,j_y+1,\uparrow} \\
c_{k_x,-,j_y+1,\uparrow}
\end{pmatrix}  \nonumber\\
&= c_{k_x,-,j_y,\uparrow}^{\dagger}c_{k_x,+,j_y+1,\uparrow}-c_{k_x,+,j_y,\uparrow}^{\dagger}c_{k_x,-,j_y+1,\uparrow},
\end{align}

\begin{align}
&C_{k_x,j_y,\uparrow}^{\dagger}(\sigma_{+}-\sigma_{-})C_{k_x,j_y+1,\uparrow} \nonumber\\
&=
\begin{pmatrix}
c_{k_x,+,j_y,\uparrow}^{\dagger} &
c_{k_x,-,j_y,\uparrow}^{\dagger}
\end{pmatrix}
\begin{pmatrix}
0 & 1 \\
-1 & 0
\end{pmatrix}
\begin{pmatrix}
c_{k_x,+,j_y+1,\uparrow} \\
c_{k_x,-,j_y+1,\uparrow}
\end{pmatrix} =\begin{pmatrix}
-c_{k_x,-,j_y,\uparrow}^{\dagger} &
c_{k_x,+,j_y,\uparrow}^{\dagger}
\end{pmatrix}
\begin{pmatrix}
c_{k_x,+,j_y+1,\uparrow} \\
c_{k_x,-,j_y+1,\uparrow}
\end{pmatrix}  \nonumber\\
&= -c_{k_x,-,j_y,\uparrow}^{\dagger}c_{k_x,+,j_y+1,\uparrow}+c_{k_x,+,j_y,\uparrow}^{\dagger}c_{k_x,-,j_y+1,\uparrow} 
= c_{k_x,+,j_y,\uparrow}^{\dagger}c_{k_x,-,j_y+1,\uparrow}-c_{k_x,-,j_y,\uparrow}^{\dagger}c_{k_x,+,j_y+1,\uparrow} \nonumber\\
&=(c_{k_x,-,j_y,\uparrow}^{\dagger}c_{k_x,+,j_y+1,\uparrow}-c_{k_x,+,j_y,\uparrow}^{\dagger}c_{k_x,-,j_y+1,\uparrow})^{\dagger}.
\end{align}
Therefore, we have
\begin{align}
&~~\sum_{{\bf k},s}C_{{\bf k},\uparrow}^{\dagger}\left[h_{\uparrow}({\bf k})\right]C_{{\bf k},\uparrow}\nonumber\\ 
&\!=\!\!-\!D\sum_{k_x,j_y}\!\left[\! A_{0}^{2} \!+\! \frac{2}{a_x^2}\!+\!\frac{2}{a_y^2}\!-\!\frac{2}{a_x^2}\cos(k_xa_x)\!\right]\!C_{k_x,j_y,\uparrow}^{\dagger}\sigma_{0}C_{k_x,j_y,\uparrow}
\!+\! \frac{D}{a_y^2}\sum_{k_x,j_y}\!\left[\!C_{k_x,j_y,\uparrow}^{\dagger}\sigma_{0}C_{k_x,j_y+1,\uparrow}\!+\! {\rm h.c.}\!\right]\! \nonumber\\
&\!+\!\sum_{k_x,j_y}\!\left[\!M\!-\!BA_{0}^{2}\!-\!\frac{2B}{a_x^2}\!-\!\frac{2B}{a_y^2}\!+\!\frac{2B}{a_x^2}\cos(k_xa_x) \!-\! \frac{A_{0}^{2}A^{2}\sin\varphi}{\hbar\omega}\!\right] \!C_{k_x,j_y,\uparrow}^{\dagger}\sigma_{z}C_{k_x,j_y,\uparrow} \!+\! \frac{B}{a_y^2}\sum_{k_x,j_y}\!\left[\!C_{k_x,j_y,\uparrow}^{\dagger}\!\sigma_{z}\!C_{k_x,j_y+1,\uparrow}\!+\! {\rm h.c.}\!\right]\! \nonumber\\
&\!+\!\left(\!1\!-\! \frac{2A_{0}^{2}B\sin\varphi}{\hbar\omega}\!\right)\frac{A}{a_x}\sum_{k_x,j_y}\sin(k_xa_x) \!\left[\!C_{k_x,j_y,\uparrow}^{\dagger}(\sigma_{+}+\sigma_{-})C_{k_x,j_y,\uparrow} \!\right]\! \nonumber\\
&\!+\!\left(\!1\!-\! \frac{2A_{0}^{2}B\sin\varphi}{\hbar\omega}\!\right) \frac{A}{2a_y}
\sum_{k_x,j_y}\!\left[\!C_{k_x,j_y,\uparrow}^{\dagger}(\sigma_{-}-\sigma_{+})C_{k_x,j_y+1,\uparrow} \!+\! {\rm h.c.}\!\right].
\end{align}

\begin{align}
&~~\sum_{{\bf k},s}C_{{\bf k},\downarrow}^{\dagger}\left[h_{\downarrow}({\bf k})\right]C_{{\bf k},\downarrow}\nonumber\\ 
&\!=\!\!-\!D\sum_{k_x,j_y}\!\left[\! A_{0}^{2} \!+\! \frac{2}{a_x^2}\!+\!\frac{2}{a_y^2}\!-\!\frac{2}{a_x^2}\cos(k_xa_x)\!\right]\!C_{k_x,j_y,\uparrow}^{\dagger}\sigma_{0}C_{k_x,j_y,\uparrow} 
\!+\! \frac{D}{a_y^2}\sum_{k_x,j_y}\!\left[\!C_{k_x,j_y,\uparrow}^{\dagger}\sigma_{0}C_{k_x,j_y+1,\uparrow}\!+\! {\rm h.c.}\!\right]\! \nonumber\\
&\!+\!\sum_{k_x,j_y}\!\left[\!M\!-\!BA_{0}^{2}\!-\!\frac{2B}{a_x^2}\!-\!\frac{2B}{a_y^2}\!+\!\frac{2B}{a_x^2}\cos(k_xa_x) \!+\! \frac{A_{0}^{2}A^{2}\sin\varphi}{\hbar\omega}\!\right] \!C_{k_x,j_y,\uparrow}^{\dagger}\sigma_{z}C_{k_x,j_y,\uparrow} \!+\! \frac{B}{a_y^2}\sum_{k_x,j_y}\!\left[\!C_{k_x,j_y,\uparrow}^{\dagger}\!\sigma_{z}\!C_{k_x,j_y+1,\uparrow}\!+\! {\rm h.c.}\!\right]\! \nonumber\\
&\!-\!\left(\!1\!+\! \frac{2A_{0}^{2}B\sin\varphi}{\hbar\omega}\!\right)\frac{A}{a_x}\sum_{k_x,j_y}\sin(k_xa_x) \!\left[\!C_{k_x,j_y,\uparrow}^{\dagger}(\sigma_{+}+\sigma_{-})C_{k_x,j_y,\uparrow} \!\right]\! \nonumber\\
&\!-\!\left(\!1\!+\! \frac{2A_{0}^{2}B\sin\varphi}{\hbar\omega}\!\right) \frac{A}{2a_y}\sum_{k_x,j_y}\!\left[\!C_{k_x,j_y,\uparrow}^{\dagger}(\sigma_{+}-\sigma_{-})C_{k_x,j_y+1,\uparrow} \!+\! {\rm h.c.}\!\right].
\end{align}

Therefore, the tight-binding Hamiltonian under $x$ periodic boundary conditions (PBCs) and $y$ open boundary conditions (OBCs) in the basis $(C_{k_x,1,\uparrow}, C_{k_x,1,\downarrow}, C_{k_x,2,\uparrow}, C_{k_x,2,\downarrow},\cdots, C_{k_x,N_y,\uparrow}, C_{k_x,N_y,\downarrow})^{T}$ is given by 
\begin{align}
{\cal H}_{y-OBC}^{(F)}(k_x)=\begin{pmatrix}
h & T & 0 & \cdots & 0 \\
T^{\dagger} & h & T & \cdots & 0 \\
0 & T^{\dagger} & h & \ddots & \vdots \\
\vdots & \ddots & \ddots & \ddots & T \\
0& \cdots & 0 & T^{\dagger} & h
\end{pmatrix}_{(4N_y)\times (4N_y)},
\end{align} where
\begin{align}
&h\!=\!\begin{pmatrix}\!
h_{\uparrow} & 0 \\
0 & h_{\downarrow}
\!\end{pmatrix},\\
&h_{\uparrow}\!=\!E_{0}\sigma_0 \!+\! \Delta_{\uparrow}\sigma_z \!+\! \Gamma_{\uparrow}(\!\sigma_{+}\!+\!\sigma_{-}\!)\!=\!E_{0}\sigma_0 \!+\! \Delta_{\uparrow}\sigma_z \!+\! \Gamma_{\uparrow}\sigma_{x},~~h_{\downarrow}\!=\!E_{0}\sigma_0 \!+\! \Delta_{\downarrow}\sigma_z \!+\! \Gamma_{\downarrow}(\!\sigma_{+}\!+\!\sigma_{-}\!)\!=\!E_{0}\sigma_0 \!+\! \Delta_{\downarrow}\sigma_z \!+\! \Gamma_{\downarrow}\sigma_{x},\\
&E_{0}\!=\!-\!D\!\left[\! A_{0}^{2} \!+\! \frac{2}{a_x^2}\!+\!\frac{2}{a_y^2}\!-\!\frac{2}{a_x^2}\cos(k_xa_x)\!\right]\!=\!-DA_{0}^{2} \!-\! \frac{2D}{a_x^2}\!-\!\frac{2D}{a_y^2}\!+\!\frac{2D}{a_x^2}\cos(k_xa_x), \\
&\Delta_{\uparrow}\!=\!M\!-\!BA_{0}^{2}\!-\!\frac{2B}{a_x^2}\!-\!\frac{2B}{a_y^2}\!+\!\frac{2B}{a_x^2}\cos(k_xa_x) \!-\! \frac{A_{0}^{2}A^{2}\sin\varphi}{\hbar\omega},~~\Delta_{\downarrow}\!=\!M\!-\!BA_{0}^{2}\!-\!\frac{2B}{a_x^2}\!-\!\frac{2B}{a_y^2}\!+\!\frac{2B}{a_x^2}\cos(k_xa_x) \!+\! \frac{A_{0}^{2}A^{2}\sin\varphi}{\hbar\omega},\\
&\Gamma_{\uparrow}\!=\!\left(\!1\!-\! \frac{2A_{0}^{2}B\sin\varphi}{\hbar\omega}\!\right)\frac{A}{a_x}\sin(k_xa_x),~~\Gamma_{\downarrow}\!=\!-\left(\!1\!+\! \frac{2A_{0}^{2}B\sin\varphi}{\hbar\omega}\!\right)\frac{A}{a_x}\sin(k_xa_x),\\
&T\!=\!\begin{pmatrix}\!
T_{\uparrow} & 0 \\
0 & T_{\downarrow}
\!\end{pmatrix}\!,~~T^{\dagger}
\!=\!\begin{pmatrix}\!
T_{\uparrow}^{\dagger} & 0 \\
0 & T_{\downarrow}^{\dagger}
\!\end{pmatrix}\!,\\
&T_{\uparrow}\!=\!
\frac{D}{a_y^2}\sigma_0 \!+\! \frac{B}{a_y^2}\sigma_z \!+\! \left(\!1\!-\! \frac{2A_{0}^{2}B\sin\varphi}{\hbar\omega}\!\right) \frac{A}{2a_y}(\sigma_{-}-\sigma_{+}),~~T^{\dagger}_{\uparrow}\!=\!
\frac{D}{a_y^2}\sigma_0 \!+\! \frac{B}{a_y^2}\sigma_z \!+\! \left(\!1\!-\! \frac{2A_{0}^{2}B\sin\varphi}{\hbar\omega}\!\right) \frac{A}{2a_y}(\sigma_{+}-\sigma_{-}),\\
&T_{\downarrow}\!=\!
\frac{D}{a_y^2}\sigma_0 \!+\! \frac{B}{a_y^2}\sigma_z \!-\! \left(\!1\!+\! \frac{2A_{0}^{2}B\sin\varphi}{\hbar\omega}\!\right) \frac{A}{2a_y}(\sigma_{-}-\sigma_{+}),~~T^{\dagger}_{\downarrow}\!=\!
\frac{D}{a_y^2}\sigma_0 \!+\! \frac{B}{a_y^2}\sigma_z \!-\! \left(\!1\!+\! \frac{2A_{0}^{2}B\sin\varphi}{\hbar\omega}\!\right) \frac{A}{2a_y}(\sigma_{+}-\sigma_{-}),
\end{align} where we use $\sigma_{+}\!+\!\sigma_{-}=\sigma_{x}$, and $\sigma_{\pm}=(\sigma_{x}\pm i\sigma_{y})/2$ are the spin-raising and spin-lowering operators, i.e.,
\begin{align}
\sigma_{+} \!=\!
\begin{pmatrix}
    0 & 1 \\
    0 & 0
\end{pmatrix},~~
\sigma_{-} \!=\!
\begin{pmatrix}
    0 & 0 \\
    1 & 0
\end{pmatrix}.
\end{align}

In addition, we can also get the tight-binding Hamiltonian with pseudo-spin-$\uparrow$ under $x$-PBCs and $y$-OBCs in the basis $(C_{k_x,1,\uparrow}, C_{k_x,2,\uparrow},\cdots, C_{k_x,N_y,\uparrow})^{T}$ as
\begin{align}
{\cal H}_{y-OBC}^{(F)\uparrow}(k_x)=\begin{pmatrix}
h_{\uparrow} & T_{\uparrow} & 0 & \cdots & 0 \\
T_{\uparrow}^{\dagger} & h_{\uparrow} & T_{\uparrow} & \cdots & 0 \\
0 & T_{\uparrow}^{\dagger} & h_{\uparrow} & \ddots & \vdots \\
\vdots & \ddots & \ddots & \ddots & T_{\uparrow} \\
0& \cdots & 0 & T_{\uparrow}^{\dagger} & h_{\uparrow}
\end{pmatrix}_{(2N_y)\times (2N_y)},
\end{align} where
\begin{align}
&h_{\uparrow}\!=\!
E_{0}\sigma_0 \!+\! \Delta_{\uparrow}\sigma_z \!+\! \Gamma_{\uparrow}(\!\sigma_{+}\!+\!\sigma_{-}\!)
\!=\!
E_{0}\sigma_0 \!+\! \Delta_{\uparrow}\sigma_z \!+\! \Gamma_{\uparrow}\sigma_{x},\\
&T_{\uparrow}\!=\!
\frac{D}{a_y^2}\sigma_0 \!+\! \frac{B}{a_y^2}\sigma_z \!+\! \left(\!1\!-\! \frac{2A_{0}^{2}B\sin\varphi}{\hbar\omega}\!\right) \frac{A}{2a_y}(\sigma_{-}-\sigma_{+}),~~T^{\dagger}_{\uparrow}\!=\!
\frac{D}{a_y^2}\sigma_0 \!+\! \frac{B}{a_y^2}\sigma_z \!+\! \left(\!1\!-\! \frac{2A_{0}^{2}B\sin\varphi}{\hbar\omega}\!\right) \frac{A}{2a_y}(\sigma_{+}-\sigma_{-}).
\end{align}

Further, we can get the tight-binding Hamiltonian with pseudo-spin-$\downarrow$ under $x$-PBCs and $y$-OBCs in the basis $(C_{k_x,1,\downarrow}, C_{k_x,2,\downarrow},\cdots, C_{k_x,N_y,\downarrow})^{T}$ as
\begin{align}
{\cal H}_{y-OBC}^{(F)\downarrow}(k_x)=\begin{pmatrix}
h_{\downarrow} & T_{\downarrow} & 0 & \cdots & 0 \\
T_{\downarrow}^{\dagger} & h_{\downarrow} & T_{\downarrow} & \cdots & 0 \\
0 & T_{\downarrow}^{\dagger} & h_{\downarrow} & \ddots & \vdots \\
\vdots & \ddots & \ddots & \ddots & T_{\downarrow} \\
0& \cdots & 0 & T_{\downarrow}^{\dagger} & h_{\downarrow}
\end{pmatrix}_{(2N_y)\times (2N_y)},
\end{align} where
\begin{align}
&h_{\downarrow}\!=\!
E_{0}\sigma_0 \!+\! \Delta_{\downarrow}\sigma_z \!+\! \Gamma_{\downarrow}(\!\sigma_{+}\!+\!\sigma_{-}\!)
\!=\!
E_{0}\sigma_0 \!+\! \Delta_{\downarrow}\sigma_z \!+\! \Gamma_{\downarrow}\sigma_{x},\\
&T_{\downarrow}\!=\!
\frac{D}{a_y^2}\sigma_0 \!+\! \frac{B}{a_y^2}\sigma_z \!-\! \left(\!1\!+\! \frac{2A_{0}^{2}B\sin\varphi}{\hbar\omega}\!\right) \frac{A}{2a_y}(\sigma_{-}-\sigma_{+}),~~T^{\dagger}_{\downarrow}\!=\!
\frac{D}{a_y^2}\sigma_0 \!+\! \frac{B}{a_y^2}\sigma_z \!-\! \left(\!1\!+\! \frac{2A_{0}^{2}B\sin\varphi}{\hbar\omega}\!\right) \frac{A}{2a_y}(\sigma_{+}-\sigma_{-}).
\end{align}

\section{Tight-binding model under the open boundary condition along the $x$ direction and periodic boundary conditions along the $y$ direction}\label{D}

The motivation for Appendix \ref{D} is to derive the analytical expression for the tight-binding model Hamiltonian under the open boundary condition along the $x$ direction and periodic boundary conditions along the $y$ direction as a comparison to Appendix \ref{C}.

In a lattice, one makes the following replacements~\cite{shen2012topological}:
\begin{align}
&k_{j}\rightarrow\frac{1}{a_j}\sin(k_{j}a_{j}),\\
&k_{j}^{2}\rightarrow\frac{2}{a_j^2}[1-\cos(k_{j}a_{j})],
\end{align} where $j=x,y,z$, $a_j$ is the lattice constant along the $j$ direction, $\sin(k_{j}a_{j})=\frac{e^{ik_{j}a_{j}}-e^{-ik_{j}a_{j}}}{2i}$, and $\cos(k_{j}a_{j})=\frac{e^{ik_{j}a_{j}}+e^{-ik_{j}a_{j}}}{2}$,
\begin{align}
&k^{2}=k_{x}^{2}+k_{y}^{2}\rightarrow\frac{2}{a_x^2}[1-\cos(k_{x}a_{x})] + \frac{2}{a_y^2}[1-\cos(k_{y}a_{y})].
\end{align}
In this way, the hopping terms in the lattice model only exist between the nearest-neighbor sites.
With this mapping, one obtains the following tight-binding model with high-frequency pumping in the basis $(c_{{\bf k},+,\uparrow}, c_{{\bf k},-,\uparrow}, c_{{\bf k},+,\downarrow}, c_{{\bf k},-,\downarrow})^{T}$ as 
\begin{align}\label{eq:HFfilm_supp}
\!{\cal H}^{(F)}({\bf k})\!=\! \left(
  \begin{array}{cccc}
    h_{\uparrow}({\bf k}) & 0  \\
    0 & h_{\downarrow}({\bf k})
  \end{array}
\right),
\end{align}
where
\begin{align}
h_{\uparrow}({\bf k})
&\!=\! \!-\! D\left\{A_{0}^{2} \!+\! \frac{2}{a_x^2}[1\!-\!\cos(k_{x}a_{x})] \!+\! \frac{2}{a_y^2}[1\!-\!\cos(k_{y}a_{y})] \right\} \sigma_0 \nonumber\\
&\!+\! \left\{M \!-\! B\left[A_{0}^{2} \!+\! \frac{2}{a_x^2}[1\!-\!\cos(k_{x}a_{x})] \!+\! \frac{2}{a_y^2}[1\!-\!\cos(k_{y}a_{y})] \right] \!-\! \frac{A_{0}^{2}A^{2}\sin\varphi}{\hbar\omega} \right\}\sigma_z \nonumber\\
&\!+\! \left(\!1\!-\! \frac{2A_{0}^{2}B\sin\varphi}{\hbar\omega}\!\right)A\left[\frac{1}{a_x}\sin(k_{x}a_{x}) \!-\! \frac{i}{a_y}\sin(k_{y}a_{y}) \right]\sigma_{+}\nonumber\\
&\!+\! \left(\!1\!-\! \frac{2A_{0}^{2}B\sin\varphi}{\hbar\omega}\!\right)A \left[\frac{1}{a_x}\sin(k_{x}a_{x}) \!+\! \frac{i}{a_y}\sin(k_{y}a_{y}) \right]\sigma_{-}, 
\end{align}
\begin{align}
h_{\downarrow}({\bf k})
&\!=\! \!-\! D\left\{A_{0}^{2} \!+\! \frac{2}{a_x^2}[1\!-\!\cos(k_{x}a_{x})] \!+\! \frac{2}{a_y^2}[1\!-\!\cos(k_{y}a_{y})] \right\} \sigma_0  \nonumber\\
&\!+\! \left\{M \!-\! B\left[A_{0}^{2} \!+\! \frac{2}{a_x^2}[1\!-\!\cos(k_{x}a_{x})] \!+\! \frac{2}{a_y^2}[1\!-\!\cos(k_{y}a_{y})] \right] \!+\! \frac{A_{0}^{2}A^{2}\sin\varphi}{\hbar\omega} \right\}\sigma_z \nonumber\\
&\!-\! \left(\!1\!+\! \frac{2A_{0}^{2}B\sin\varphi}{\hbar\omega}\!\right)A\left[\frac{1}{a_x}\sin(k_{x}a_{x}) \!+\! \frac{i}{a_y}\sin(k_{y}a_{y}) \right]\sigma_{+}\nonumber\\
&\!-\! \left(\!1\!+\! \frac{2A_{0}^{2}B\sin\varphi}{\hbar\omega}\!\right)A \left[\frac{1}{a_x}\sin(k_{x}a_{x}) \!-\! \frac{i}{a_y}\sin(k_{y}a_{y}) \right]\sigma_{-}.
\end{align}

Performing the Fourier transformation, one obtains
\begin{align}
c_{{\bf k},s,\sigma}&=\frac{1}{\sqrt{N_{x}}}\sum_{j_x,k_y}^{N_x}e^{-ik_{x}j_{x}a_{x}}c_{j_x,k_y,s,\sigma},\\
c_{{\bf k},s,\sigma}^{\dagger}&=\frac{1}{\sqrt{N_{x}}}\sum_{j_x,k_y}^{N_x}e^{ik_{x}j_{x}a_{x}}c_{j_x,k_y,s,\sigma}^{\dagger}.
\end{align}
\begin{align}
C_{{\bf k},\sigma}&=\begin{pmatrix}
c_{{\bf k},+,\sigma} \\
c_{{\bf k},-,\sigma}
\end{pmatrix}=\frac{1}{\sqrt{N_{x}}}\sum_{j_x,k_y}^{N_x}e^{-ik_{x}j_{x}a_{x}}C_{j_x,k_y,\sigma},\\
C_{{\bf k},\sigma}^{\dagger}&=\begin{pmatrix}
c_{{\bf k},+,\sigma}^{\dagger} &
c_{{\bf k},-,\sigma}^{\dagger}
\end{pmatrix}=\frac{1}{\sqrt{N_{x}}}\sum_{j_x,k_y}^{N_x}e^{ik_{x}j_{x}a_{x}}C_{j_x,k_y,\sigma}^{\dagger}.
\end{align} 

\begin{align}
&~~\sum_{{\bf k},s}C_{{\bf k},\uparrow}^{\dagger}\left[h_{\uparrow}({\bf k})\right]C_{{\bf k},\uparrow}\nonumber\\ 
&\!=\!\!-\!D\sum_{j_x,k_y}\!\left[\! A_{0}^{2} \!+\! \frac{2}{a_x^2}\!+\!\frac{2}{a_y^2}\!-\!\frac{2}{a_y^2}\cos(k_ya_y)\!\right]\!C_{j_x,k_y,\uparrow}^{\dagger}\sigma_{0}C_{j_x,k_y,\uparrow} \!+\! \frac{D}{a_x^2}\sum_{j_x,k_y}\!\left[\!C_{j_x,k_y,\uparrow}^{\dagger}\sigma_{0}C_{j_x+1,k_y,\uparrow}\!+\! {\rm h.c.}\!\right]\! \nonumber\\
&\!+\!\sum_{j_x,k_y}\!\left[\!M\!-\!BA_{0}^{2}\!-\!\frac{2B}{a_x^2}\!-\!\frac{2B}{a_y^2}\!+\!\frac{2B}{a_y^2}\cos(k_ya_y) \!-\! \frac{A_{0}^{2}A^{2}\sin\varphi}{\hbar\omega}\!\right] \!C_{j_x,k_y,\uparrow}^{\dagger}\sigma_{z}C_{j_x,k_y,\uparrow} \!+\! \frac{B}{a_x^2}\sum_{j_x,k_y}\!\left[\!C_{j_x,k_y,\uparrow}^{\dagger}\!\sigma_{z}\!C_{j_x+1,k_y,\uparrow}\!+\! {\rm h.c.}\!\right]\! \nonumber\\
&\!+\!\left(\!1\!-\! \frac{2A_{0}^{2}B\sin\varphi}{\hbar\omega}\!\right)\frac{A}{a_y}\sum_{j_x,k_y}\sin(k_ya_y) \!\left[\!C_{j_x,k_y,\uparrow}^{\dagger}(-i\sigma_{+}+i\sigma_{-})C_{j_x,k_y,\uparrow} \!\right]\! \nonumber\\
&\!+\!\left(\!1\!-\! \frac{2A_{0}^{2}B\sin\varphi}{\hbar\omega}\!\right) \frac{A}{2a_x}\sum_{j_x,k_y}\!\left[\!-iC_{j_x,k_y,\uparrow}^{\dagger}(\sigma_{+}+\sigma_{-})C_{j_x+1,k_y,\uparrow} \!+\! {\rm h.c.}\!\right].
\end{align}

\begin{align}
&~~\sum_{{\bf k},s}C_{{\bf k},\downarrow}^{\dagger}\left[h_{\downarrow}({\bf k})\right]C_{{\bf k},\downarrow}\nonumber\\ 
&\!=\!\!-\!D\sum_{j_x,k_y}\!\left[\! A_{0}^{2} \!+\! \frac{2}{a_x^2}\!+\!\frac{2}{a_y^2}\!-\!\frac{2}{a_y^2}\cos(k_ya_y)\!\right]\!C_{j_x,k_y,\uparrow}^{\dagger}\sigma_{0}C_{j_x,k_y,\uparrow} \!+\! \frac{D}{a_x^2}\sum_{j_x,k_y}\!\left[\!C_{j_x,k_y,\uparrow}^{\dagger}\sigma_{0}C_{j_x+1,k_y,\uparrow}\!+\! {\rm h.c.}\!\right]\! \nonumber\\
&\!+\!\sum_{j_x,k_y}\!\left[\!M\!-\!BA_{0}^{2}\!-\!\frac{2B}{a_x^2}\!-\!\frac{2B}{a_y^2}\!+\!\frac{2B}{a_y^2}\cos(k_ya_y) \!+\! \frac{A_{0}^{2}A^{2}\sin\varphi}{\hbar\omega}\!\right] \!C_{j_x,k_y,\uparrow}^{\dagger}\sigma_{z}C_{j_x,k_y,\uparrow} \!+\! \frac{B}{a_x^2}\sum_{j_x,k_y}\!\left[\!C_{j_x,k_y,\uparrow}^{\dagger}\!\sigma_{z}\!C_{j_x+1,k_y,\uparrow}\!+\! {\rm h.c.}\!\right]\! \nonumber\\
&\!-\!\left(\!1\!+\! \frac{2A_{0}^{2}B\sin\varphi}{\hbar\omega}\!\right)\frac{A}{a_y}\sum_{j_x,k_y}\sin(k_ya_y) \!\left[\!C_{j_x,k_y,\uparrow}^{\dagger}(i\sigma_{+}-i\sigma_{-})C_{j_x,k_y,\uparrow} \!\right]\! \nonumber\\
&\!-\!\left(\!1\!+\! \frac{2A_{0}^{2}B\sin\varphi}{\hbar\omega}\!\right) \frac{A}{2a_x} \sum_{j_x,k_y}\!\left[\!-iC_{j_x,k_y,\uparrow}^{\dagger}(\sigma_{+}+\sigma_{-})C_{j_x+1,k_y,\uparrow} \!+\! {\rm h.c.}\!\right].
\end{align}

Therefore, the tight-binding Hamiltonian under $y$-PBCs and $x$-OBCs in the basis $(C_{1,k_y,\uparrow}, C_{1,k_y,\downarrow}, C_{2,k_y,\uparrow}, C_{2,k_y,\downarrow},\cdots, C_{N_x,k_y,\uparrow}, C_{N_x,k_y,\downarrow})^{T}$ is given by 
\begin{align}
{\cal H}_{x-OBC}^{(F)}(k_y)=\begin{pmatrix}
h & T & 0 & \cdots & 0 \\
T^{\dagger} & h & T & \cdots & 0 \\
0 & T^{\dagger} & h & \ddots & \vdots \\
\vdots & \ddots & \ddots & \ddots & T \\
0& \cdots & 0 & T^{\dagger} & h
\end{pmatrix}_{(4N_x)\times (4N_x)},
\end{align} where
\begin{align}
&h\!=\!\begin{pmatrix}\!
h_{\uparrow} & 0 \\
0 & h_{\downarrow}
\!\end{pmatrix},\\
&h_{\uparrow}\!=\!
E_{0}\sigma_0 \!+\! \Delta_{\uparrow}\sigma_z \!+\! \Gamma_{\uparrow}(\!-i\sigma_{+}\!+\!i\sigma_{-}\!)\!=\!
E_{0}\sigma_0 \!+\! \Delta_{\uparrow}\sigma_z \!+\! \Gamma_{\uparrow}\sigma_{y}, ~~h_{\downarrow}\!=\!
E_{0}\sigma_0 \!+\! \Delta_{\downarrow}\sigma_z \!+\! \Gamma_{\downarrow}(\!i\sigma_{+}\!-\!i\sigma_{-}\!)\!=\!
E_{0}\sigma_0 \!+\! \Delta_{\downarrow}\sigma_z \!-\! \Gamma_{\downarrow}\sigma_{y}, \\
&E_{0}=\!\!-DA_{0}^{2} \!-\! \frac{2D}{a_x^2}\!-\!\frac{2D}{a_y^2}\!+\!\frac{2D}{a_y^2}\cos(k_ya_y), \\
&\Delta_{\uparrow}=M\!-\!BA_{0}^{2}\!-\!\frac{2B}{a_x^2}\!-\!\frac{2B}{a_y^2}\!+\!\frac{2B}{a_y^2}\cos(k_ya_y) \!-\! \frac{A_{0}^{2}A^{2}\sin\varphi}{\hbar\omega},~~\Delta_{\downarrow}=M\!-\!BA_{0}^{2}\!-\!\frac{2B}{a_x^2}\!-\!\frac{2B}{a_y^2}\!+\!\frac{2B}{a_y^2}\cos(k_ya_y) \!+\! \frac{A_{0}^{2}A^{2}\sin\varphi}{\hbar\omega},\\
&\Gamma_{\uparrow}=\left(\!1\!-\! \frac{2A_{0}^{2}B\sin\varphi}{\hbar\omega}\!\right)\frac{A}{a_y}\sin(k_ya_y),~~\Gamma_{\downarrow}=-\left(\!1\!+\! \frac{2A_{0}^{2}B\sin\varphi}{\hbar\omega}\!\right)\frac{A}{a_y}\sin(k_ya_y),\\
&T\!=\!\begin{pmatrix}\!
T_{\uparrow} & 0 \\
0 & T_{\downarrow}
\!\end{pmatrix}\!,~~T^{\dagger}
\!=\!\begin{pmatrix}\!
T_{\uparrow}^{\dagger} & 0 \\
0 & T_{\downarrow}^{\dagger}
\!\end{pmatrix}\!,\\
&T_{\uparrow}\!=\!
\frac{D}{a_x^2}\sigma_0 \!+\! \frac{B}{a_x^2}\sigma_z \!+\! \left(\!1\!-\! \frac{2A_{0}^{2}B\sin\varphi}{\hbar\omega}\!\right) \frac{(-iA)}{2a_x}(\sigma_{+}+\sigma_{-}),~~T^{\dagger}_{\uparrow}\!=\!
\frac{D}{a_x^2}\sigma_0 \!+\! \frac{B}{a_x^2}\sigma_z \!+\! \left(\!1\!-\! \frac{2A_{0}^{2}B\sin\varphi}{\hbar\omega}\!\right) \frac{iA}{2a_x}(\sigma_{+}+\sigma_{-}),\\
&T_{\downarrow}\!=\!
\frac{D}{a_x^2}\sigma_0 \!+\! \frac{B}{a_x^2}\sigma_z \!-\! \left(\!1\!+\! \frac{2A_{0}^{2}B\sin\varphi}{\hbar\omega}\!\right) \frac{(-iA)}{2a_x}(\sigma_{+}+\sigma_{-}),~~T^{\dagger}_{\downarrow}\!=\!
\frac{D}{a_x^2}\sigma_0 \!+\! \frac{B}{a_x^2}\sigma_z \!-\! \left(\!1\!+\! \frac{2A_{0}^{2}B\sin\varphi}{\hbar\omega}\!\right) \frac{iA}{2a_x}(\sigma_{+}+\sigma_{-}),
\end{align} where we use $\sigma_{+}\!+\!\sigma_{-}=\sigma_{x}$.

In addition, we can also get the tight-binding Hamiltonian with pseudo-spin-$\uparrow$ under $y$-PBCs and $x$-OBCs in the basis $(C_{1,k_y,\uparrow}, C_{2,k_y,\uparrow},\cdots, C_{N_x,k_y,\uparrow})^{T}$ as
\begin{align}
{\cal H}_{x-OBC}^{(F)\uparrow}(k_y)=\begin{pmatrix}
h_{\uparrow} & T_{\uparrow} & 0 & \cdots & 0 \\
T_{\uparrow}^{\dagger} & h_{\uparrow} & T_{\uparrow} & \cdots & 0 \\
0 & T_{\uparrow}^{\dagger} & h_{\uparrow} & \ddots & \vdots \\
\vdots & \ddots & \ddots & \ddots & T_{\uparrow} \\
0& \cdots & 0 & T_{\uparrow}^{\dagger} & h_{\uparrow}
\end{pmatrix}_{(2N_x)\times (2N_x)},
\end{align} where
\begin{align}
&h_{\uparrow}\!=\!
E_{0}\sigma_0 \!+\! \Delta_{\uparrow}\sigma_z \!+\! \Gamma_{\uparrow}(\!-i\sigma_{+}\!+\!i\sigma_{-}\!) 
\!=\!
E_{0}\sigma_0 \!+\! \Delta_{\uparrow}\sigma_z \!-\! i\Gamma_{\uparrow}(\!\sigma_{+}\!-\!\sigma_{-}\!)\!=\!
E_{0}\sigma_0 \!+\! \Delta_{\uparrow}\sigma_z \!+\! \Gamma_{\uparrow}\sigma_{y},\\
&T_{\uparrow}\!=\!
\frac{D}{a_x^2}\sigma_0 \!+\! \frac{B}{a_x^2}\sigma_z \!+\! \left(\!1\!-\! \frac{2A_{0}^{2}B\sin\varphi}{\hbar\omega}\!\right) \frac{(-iA)}{2a_x}(\sigma_{+}+\sigma_{-}),~~T^{\dagger}_{\uparrow}\!=\!
\frac{D}{a_x^2}\sigma_0 \!+\! \frac{B}{a_x^2}\sigma_z \!+\! \left(\!1\!-\! \frac{2A_{0}^{2}B\sin\varphi}{\hbar\omega}\!\right) \frac{iA}{2a_x}(\sigma_{+}+\sigma_{-}).
\end{align}

Further, we can get the tight-binding Hamiltonian with pseudo-spin-$\downarrow$ under $y$-PBCs and $x$-OBCs in the basis $(C_{1,k_y,\downarrow}, C_{2,k_y,\downarrow},\cdots, C_{N_x,k_y,\downarrow})^{T}$ as
\begin{align}
{\cal H}_{x-OBC}^{(F)\downarrow}(k_y)=\begin{pmatrix}
h_{\downarrow} & T_{\downarrow} & 0 & \cdots & 0 \\
T_{\downarrow}^{\dagger} & h_{\downarrow} & T_{\downarrow} & \cdots & 0 \\
0 & T_{\downarrow}^{\dagger} & h_{\downarrow} & \ddots & \vdots \\
\vdots & \ddots & \ddots & \ddots & T_{\downarrow} \\
0& \cdots & 0 & T_{\downarrow}^{\dagger} & h_{\downarrow}
\end{pmatrix}_{(2N_x)\times (2N_x)},
\end{align} where
\begin{align}
&h_{\downarrow}\!=\!
E_{0}\sigma_0 \!+\! \Delta_{\downarrow}\sigma_z \!+\! \Gamma_{\downarrow}(\!i\sigma_{+}\!-\!i\sigma_{-}\!) 
\!=\!
E_{0}\sigma_0 \!+\! \Delta_{\downarrow}\sigma_z \!+\! i\Gamma_{\downarrow}(\!\sigma_{+}\!-\!\sigma_{-}\!)\!=\!
E_{0}\sigma_0 \!+\! \Delta_{\downarrow}\sigma_z \!-\! \Gamma_{\downarrow}\sigma_{y},\\
&T_{\downarrow}\!=\!
\frac{D}{a_x^2}\sigma_0 \!+\! \frac{B}{a_x^2}\sigma_z \!-\! \left(\!1\!+\! \frac{2A_{0}^{2}B\sin\varphi}{\hbar\omega}\!\right) \frac{(-iA)}{2a_x}(\sigma_{+}+\sigma_{-}),~~T^{\dagger}_{\downarrow}\!=\!
\frac{D}{a_x^2}\sigma_0 \!+\! \frac{B}{a_x^2}\sigma_z \!-\! \left(\!1\!+\! \frac{2A_{0}^{2}B\sin\varphi}{\hbar\omega}\!\right) \frac{iA}{2a_x}(\sigma_{+}+\sigma_{-}).
\end{align}


\section{Probability distributions of the edge states for $E_{\sigma-}$}\label{E}

The motivation for Appendix \ref{E} is to give the probability distributions of the edge states for $E_{\sigma-}$ as supplemental material for Fig.~\ref{Fig:Vz_tb} in the main text.

\begin{figure}[htpb]
\centering
\includegraphics[width=0.48\textwidth]{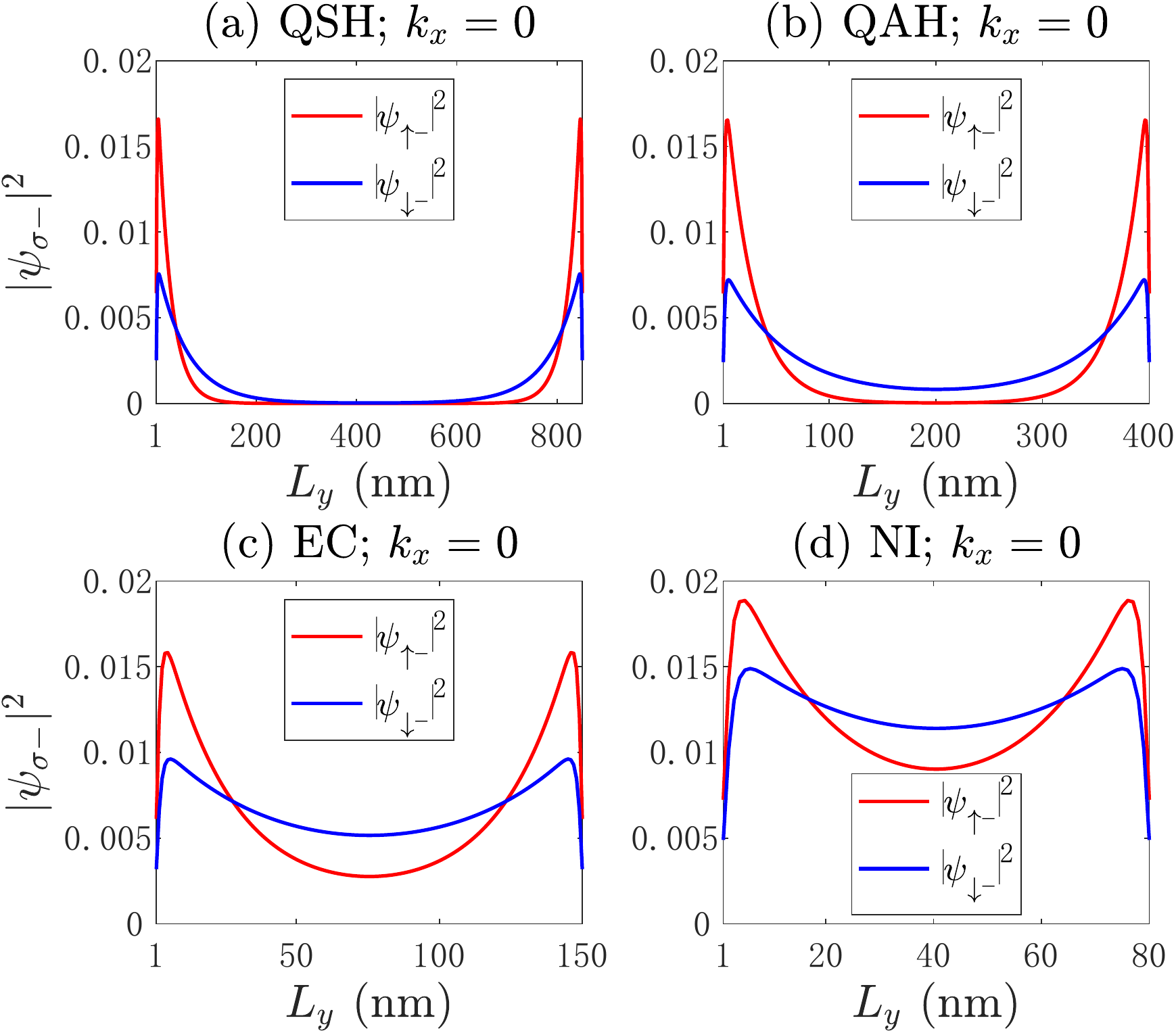}
\caption{(Color online) Probability distributions of the edge states for the Floquet Hamiltonian of the HgTe/(Hg,Cd)Te with different sizes $L_y$ under $k_{x}=0$. The other parameters are the same as those in Fig.~\ref{Fig:E_tb}.
Notice that the subscript $-$ denotes the edge states of $E_{\sigma-}$ with $\sigma=\uparrow,\downarrow$.}  \label{Fig:Vf_tb}
\end{figure}

It is indicated in Fig.~\ref{Fig:Vf_tb} that the probability distributions of the edge states for the $-$ bands $E_{\sigma-}$ have similar properties as those for the $+$ bands $E_{\sigma+}$ with changing system size $L_y$.

\end{widetext}

\end{document}